\documentclass[aps,prb,twocolumn,showpacs]{revtex4}
\usepackage{amssymb,amsmath}
\usepackage{graphicx}
\usepackage{epsfig}   
\begin{document}

\title{Topological Order, Dimerization, and Spinon Deconfinement\\
in Frustrated Spin Ladders}
\author{Eugene H. Kim,$^1$ \"O. Legeza,$^2$ and J. S\'olyom$^2$ }
\affiliation{$^1$Department of Physics, University of Windsor,
   Windsor, Ontario, Canada N9B 3P4   \\
$^2$Research Institute for Solid State Physics and Optics,
   P. O. Box 49, H-1525 Budapest, Hungary}
\date{\today}
                                                  
\begin{abstract}
We consider topological order and dimer order in several frustrated 
spin ladder models, which are related to higher dimensional models
of current interest; we also address the occurrence of fractionalized 
phases with deconfined spinon excitations in these models.
Combining results obtained with both analytic and numerical methods,
we discuss how the occurrence of dimerized or fractionalized phases
are dictated by the system's geometry.
\end{abstract}

\maketitle

%%%%%%%%%%%%%%%%%%%%%%%%%%%%%%%%%%%%%%%%%%%%%
%%%%%%%%%%%%%%%%%%%%%%%%%%%%%%%%%%%%%%%%%%%%%

\section{Introduction}

Frustrated spin systems give rise to a wealth of interesting behavior; 
hence, they have attracted considerable attention.\cite{frustrated} 
Typically, the low-lying excitations in spin systems are magnons.  
However, there has been particular interest in identifying fractionalized 
phases, where the magnon ``breaks apart" into more fundamental 
objects.  When doped, such phases would give rise to metallic states 
which do not fall within the Fermi liquid paradigm --- the elementary 
excitations in these phases would not have the quantum numbers of 
an electron. However, finding systems which exhibit fractionalized 
phases has proven to be extremely challenging.

Motivated largely by Anderson's original resonating valence bond (RVB)
suggestions for the high-$T_c$ cuprate superconductors,\cite{anderson} 
substantial effort has focussed on searching for fractionalized phases in 
models having short-range RVB ground states.\cite{2Dfraction} 
A related approach has been to consider dimer models, where only 
short-ranged valence bonds with specified dynamics are 
considered.\cite{dimer} While lacking long-range order 
in the conventional sense, i.e., lacking a local order parameter, 
these short-range valence-bond ground states have a subtle form 
of order, namely topological order.\cite{bonesteel}
It is now appreciated that a precise characterization of
fractionalized phases is via its topological order.\cite{topological}
However, whether a fractionalized phase actually occurs depends 
strongly on the system's geometry. More specifically, the system's 
geometry must allow for "liquidity" in the spectrum of states.\cite{liquidity}
Many systems with short-range RVB ground states 
do not exhibit such liquidity --- they would prefer to dimerize, 
rather than exhibit a fractionalized phase.\cite{instanton}

In this work, we address the occurrence of fractionalized and dimerized
phases in several frustrated spin ladder models.  Ladder models provide 
a unique testing ground, as powerful analytic and numerical techniques 
from one-dimensional physics can be utilized.  
Indeed, ladders models have allowed controlled calculations to investigate
topological order,\cite{white-trans,gene,fath} the occurrence of dimer 
order,\cite{oleg} and the occurrence of fractionalized excitations\cite{allen}  
in spin models. Furthermore, ladders models have allowed for controlled 
calculations demonstrating that pairing and, in particular, $d_{x^2-y^2}$ 
pairs could arise when these spin models are doped.\cite{pairing}
The models we consider are related to higher-dimensional systems of 
current interest; investigations of the one-dimensional analogs are 
particularly relevant, in light of recent work which showed that the 
mechanism giving rise to fractionalized excitations is the same in 
both one and two dimensions.\cite{dung} However,
besides being a testing ground and illustrating the types of possible
behaviors, these ladder models are interesting in their own right, as 
there are a number of materials that are well described by ladder 
models.\cite{ladderreview}

In large regions of parameter space, the models we consider have 
short-range RVB ground states. More specifically, in large regions of 
parameter space these models have ground states which are 
continuously related to the ground states of the so-called rung-singlet 
phase or the Haldane phase. [Typical configurations in these ground 
states are shown schematically in Figs.~\ref{fig:ordinaryrvb} and 
\ref{fig:compositervb} in Sec.~II.] While these phases have nearly 
identical properties, their ground states differ in a subtle way, namely 
in their topological order. With frustrating interactions, these states 
could be tuned to become degenerate, and then one has the 
necessary liquidity for the deconfinement of spinon excitations. 
However, as will be seen below, depending on the model's spatial 
symmetries, this liquidity and spinon deconfinement could be 
preempted by dimerization.

The rest of the paper is organized as follows. In Sec.~II, we describe 
the ladder models considered in this work --- the cross-coupled, 
zigzag, and diagonal ladders --- and we recall some of their known 
properties. In Sec.~III, the exact ground states of the models are 
discussed (along certain lines in parameter space). Sec.~IV contains 
a discussion of topological order in these models and some comments 
on the low-energy excitations. In Sec.~V, we analyze the models in 
the limit of weak interchain couplings, using bosonization and 
renormalization group techniques. In Sec.~VI, we present numerical 
results obtained via the density-matrix renormalization-group (DMRG) 
algorithm; we discuss these numerical results in light of the results from 
the previous sections.  Finally, in Sec.~VII we summarize and present 
some concluding remarks.

%%%%%%%%%%%%%%%%%%%%%%%%%%%%%%%%%%%%%%%%%%%%%%
%%%%%%%%%%%%%%%%%%%%%%%%%%%%%%%%%%%%%%%%%%%%%

\section{The Models}

We begin with two antiferromagnetic spin-1/2 Heisenberg chains, each 
described by the Hamiltonian
\begin{equation}
H_0^{(i)} = \sum_l J_{\|}~ 
{\bf S}^{(i)}_l \cdot {\bf S}^{(i)}_{l+1}  \qquad i=1,2
\end{equation}
where ${\bf S}^{(i)}_{l}$ is the spin operator at site $l$ on chain $i$. 
We will be interested in coupling the chains together in various ways,
such that the resulting models have different spatial symmetries.  
In this work, we will consider only antiferromagnetic couplings.
We start by coupling the chains together so that the resulting models 
are not frustrated.  We then include frustrating interactions and
investigate their influence on the properties of the models.

\begin{figure}[htb]
\scalebox{.325}{\includegraphics{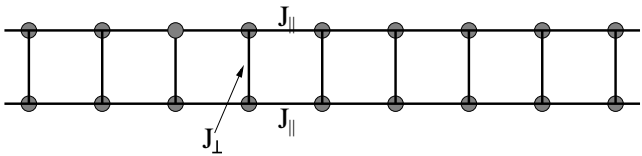}}
\caption{The ordinary ladder.}
\label{fig:ordinary}
\end{figure}

The simplest way to couple the chains together is by
\begin{equation}
H_{\bot} = \sum_l  J_{\bot}~ 
{\bf S}^{(1)}_{l} \cdot {\bf S}^{(2)}_l  \,,
\label{ordinaryladder}
\end{equation}
so that the full Hamiltonian is 
\begin{equation}
H_{\text{L}} = \sum_{i=1}^{2} H_0^{(i)} + H_{\perp}   
\label{Hordinary}
\end{equation}
which is depicted in Fig.~\ref{fig:ordinary}. We will refer to this as 
an {\sl ordinary ladder}. This model has received considerable 
attention.  Indeed, it can be thought of as a strip of a 
two-dimensional square lattice; hence, its properties have been 
investigated to give clues as to the physics occurring in the 
high-$T_c$ cuprate superconductors.\cite{toycuprate}

\begin{figure}[htb]
\scalebox{.305}{\includegraphics{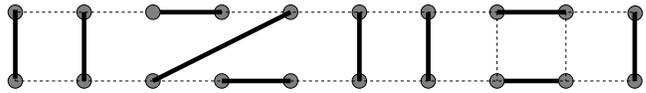}}
\caption{A typical valence-bond configuration in the rung-singlet ground 
state of the two-leg ordinary Heisenberg ladder.}
\label{fig:ordinaryrvb}
\end{figure}

When $J_{\bot}$ is large ($J_{\bot} \gg J_{\|}$), the ground state 
is essentially a product of rung singlets with a gap to the excited 
states --- the gap is due to the energy necessary to break a singlet 
bond.  When $J_{\bot} \simeq J_{\|}$, it has been shown that the 
energy gap persists; the ground state is well described by a 
short-range valence-bond state, a typical configuration of which 
is shown in Fig.~\ref{fig:ordinaryrvb}.  It has been established that 
the entire region $0 < J_{\bot} < \infty$ is, in fact, continuously 
related.\cite{wns,shelton}  As this entire region of parameter space 
is related to the regime $J_{\bot} \gg J_{\|}$ where the ground state 
is a product of rung singlets, it is often referred to as the 
{\sl rung-singlet phase}.
Incidentally, the dominance of rung-singlet bonds can be measured 
by determining their weight in the ground state 
$\rho_s = (1/N)\sum_{l=1}^{N} \langle \Psi |
S^{\phantom \dagger}_l S^{\dagger}_l | \Psi \rangle$
where $|\Psi \rangle$ is the ground-state wave function and
$S^{\phantom \dagger}_l = (1/\sqrt{2})\big[  | \! \uparrow \rangle_l^{(1)}
 | \! \downarrow \rangle_l^{(2)} - | \!\downarrow  \rangle_l^{(1)} 
| \! \uparrow \rangle_l^{(2)} \big]$. A state belongs to the rung-singlet 
phase if $\rho_s >1/4$.\cite{wang}

\begin{figure}[htb]
\scalebox{.325}{\includegraphics{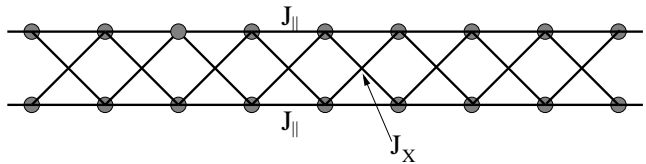}}
\caption{Ladder with diagonal couplings.}
\label{fig:cross-coupled}
\end{figure}

We also consider coupling the spins to their next-nearest neighbors 
on the opposite leg of the ladder. This coupling is described by  
\begin{equation}
H_{\text{X}} = \sum_l J_{\text{X}} \left( {\bf S}^{(1)}_l \cdot 
{\bf S}^{(2)}_{l+1} + {\bf S}^{(2)}_l \cdot {\bf S}^{(1)}_{l+1} \right) \,,
\label{J1J2}
\end{equation}
so that the full Hamiltonian is
\begin{equation}
H_{\text{C}} = \sum_{i=1}^{2} H_0^{(i)}  +  H_{\text{X}}  \,;
\label{HCCC}
\end{equation}
the resulting model is shown in Fig.~\ref{fig:cross-coupled}. 

\begin{figure}[htb]
\scalebox{.305}{\includegraphics{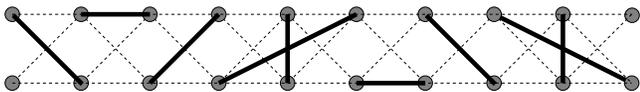}}
\caption{A typical valence-bond configuration in the ground state of the
ladder model shown in Fig.~\ref{fig:cross-coupled}.}
\label{fig:compositervb}
\end{figure}

This system is known\cite{lfs,ls} to have a gapped spectrum of spin-1 
magnons.  Furthermore, the entire regime $0 < J_{\text{X}} < \infty$ 
is continuously related to the {\sl Haldane phase}\cite{haldane} of the 
spin-1 chain. This can be understood by considering the point 
$J_{\text{X}} = J_{\|}$ --- here, the low-energy spectrum of \eqref{HCCC} 
is equivalent to that of the $S=1$ Heisenberg spin chain.\cite{timonen,xian} 
To see this, we start with the Hamiltonian of a spin-1 chain
$H = J_{\|} \sum_l {\bf S}_l \cdot {\bf S}_{l+1}$, 
which is known to be in the Haldane phase. When the spin-1 operator on 
site $l$ is represented as a sum of two spin-1/2 operators, 
${\bf S}_{l} = {\bf S}^{(1)}_{l} + {\bf S}^{(2)}_l$, one obtains \eqref{HCCC}
with $J_{\text{X}} = J_{\|}$. Since the total spin of each rung commutes with the 
Hamiltonian, the eigenstates can be classified by the total spins 
on the rungs. It has been shown that in the low-energy part of the 
spectrum all rungs are in their triplet ($S=1$) state, and hence the 
same Haldane gap appears in the ladder model as well. In this 
representation the Haldane state of the spin-1 chain can be described 
rather well by short-ranged valence bonds between neighboring rungs, 
a typical configuration of which is shown in Fig.~\ref{fig:compositervb}. 

\begin{figure}[htb]
\scalebox{.31}{\includegraphics{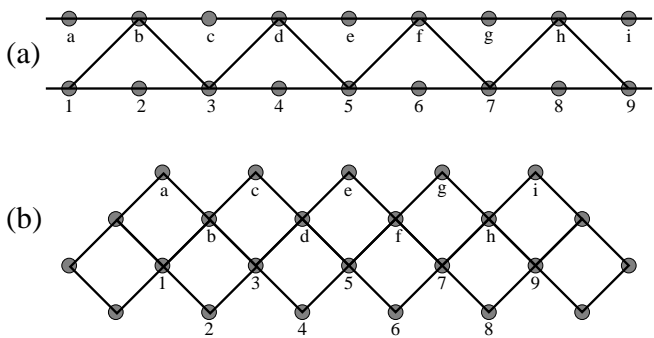}}
\caption{(a) Ladder with a diagonal coupling between every second
spin.  (b) $N_p=2$ diagonal ladder. }
\label{fig:diagonalladder}
\end{figure}

Finally, we consider the ladder model shown in 
Fig.~\ref{fig:diagonalladder}(a), where the interchain coupling is given by 
\begin{equation}
H_{\text{d}} = \sum_l  J_2~  {\bf S}^{(1)}_{2l} \cdot 
\left( {\bf S}^{(2)}_{2l-1} + {\bf S}^{(2)}_{2l+1} \right)  \,.
\label{diagonal}
\end{equation}
Previous work has established\cite{sierra} that this model is in the same 
universality class as the model in \eqref{HCCC}. 
Hence, removing half of the cross couplings in an appropriate way
from Eq.~\eqref{HCCC} does not change the universality class.
Part of our motivation for considering this model is because it is 
a representation of the ``$N_p = 2$ diagonal ladder" shown in 
Fig.~\ref{fig:diagonalladder}(b); it is a minimal model to study 
diagonal stripes, which have been observed in the high-$T_c$ cuprate 
superconducting material La$_{2-x}$Sr$_x$CuO$_4$,\cite{wakimoto}   
and the nickel oxides La$_2$NiO$_{4.125}$\cite{tranquada1} and 
La$_{1-x}$Sr$_x$NiO$_4$.\cite{tranquada2}

%%%%%%%%%%%%%%%%%%%%

As mentioned above, we are interested in the effect of frustration on 
these models.  In what follows, we analyze three models obtained by 
introducing frustrating interactions to the models described above.  
(Again, we will consider only antiferromagnetic interchain couplings 
in this work.)  Our motivation for doing so is because the resulting 
models have different spatial symmetries.  As discussed below, these 
symmetries play a crucial role in determining the properties of the 
models.

%%%%%%%%%%%%%%%%%%%%

\subsubsection{Cross-Coupled Ladder}

\begin{figure}[htb]
\scalebox{.325}{\includegraphics{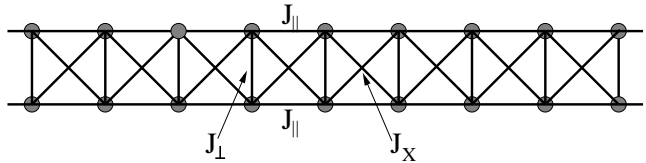}}
\caption{The {\sl cross-coupled} ladder.}
\label{fig:composite}
\end{figure}

The first model we consider has both $H_{\bot}$ and $H_{\text{X}}$ 
present simultaneously.  The resulting model with Hamiltonian
\begin{equation}
H_{\text{CC}} = \sum_{i=1}^{2} H_0^{(i)} + H_{\bot} +  H_{\text{X}} 
\label{Hcross}
\end{equation}
is shown in Fig.~\ref{fig:composite}. For the rest of this work, we will refer 
to this model as a
\emph{cross-coupled} ladder.
Notice that this spin model is invariant under translation by a single 
site; it is also invariant if the legs of the ladder are interchanged: 
leg 1 $\leftrightarrow$ leg 2. Part of our motivation for the study of 
this model is its relationship to the two-dimensional model with 
nearest-neighbor and next-nearest-neighbor exchange couplings, often 
referred to as the $J_1 - J_2$ model --- \eqref{Hcross} is a 
one-dimensional strip of this model. Furthermore, \eqref{Hcross} 
can be thought of as a chain of edge-sharing tetrahedra.\cite{martin} 
As there are a number of materials described by spin models on 
corner-sharing tetrahedra, i.e., pyrochlores,\cite{pyrochlore} it 
is not unreasonable that materials with edge-sharing tetrahedra can be 
realized.

The model in \eqref{Hcross} has been investigated in a number of works,
but its phase diagram and properties are still under debate.  When $J_{\bot}$ 
is the dominant interchain coupling, the system is continuously related 
to the ordinary ladder (with $J_{\text{X}}=0$); hence, the system is in the 
rung-singlet phase.  On the other hand, when $J_{\text{X}}$ is the dominant 
interchain coupling, the system is continuously related to Eq.~\eqref{HCCC}, 
and the system is in the Haldane phase. For weak interchain couplings, 
previous analytic treatments\cite{gene} suggested a first-order transition 
between the rung-singlet and Haldane phases when 
$J_{\bot} \simeq 2J_{\text{X}}$.  Numerical results on the 
model\cite{weihong,fath} were consistent with a first-order transition for 
both weak and strong interchain couplings. Recent numerical work, however, 
has suggested the transition is actually continuous when the interchain coupling 
is weak, becoming first-order only for stronger interchain coupling.\cite{wang,hung}  
Furthermore, recent analytic work\cite{oleg} has argued there is a 
spontaneously dimerized phase in between the rung-singlet and Haldane phases, 
rather than a direct transition between the two phases as observed in previous works.

%%%%%%%%%%%%%%%%%%%%

\subsubsection{Zigzag Ladder}

\begin{figure}[htb]
\scalebox{.325}{\includegraphics{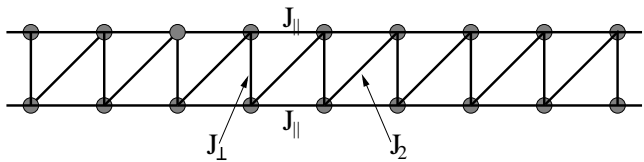}}
\caption{The zigzag ladder model. }
\label{fig:zigzag}
\end{figure}

The second ladder model we consider contains again both rung and
diagonal couplings, but only half the diagonal couplings of the
cross-coupled ladder are present, as shown in Fig.~\ref{fig:zigzag}.
This model is often referred to as a {\sl zigzag ladder}; its 
Hamiltonian is
\begin{equation}
H_{\text{Z}} = \sum_{i=1}^2 H_0^{(i)} + H_{\bot} + H_{\text{z}}  \,,
\label{triangular}
\end{equation}
where 
\begin{equation}
H_{\text{z}} = \sum_l J_2~ {\bf S}^{(1)}_l \cdot {\bf S}^{(2)}_{l+1}  \,.
\end{equation}
This spin model is invariant under translation by a single site.  
However, unlike the cross-coupled ladder which is invariant under 
the interchange of the legs, this model lacks that symmetry.
Part of the motivation for considering this model comes from the 
two-dimensional triangular lattice --- \eqref{triangular} is a 
one-dimensional strip of the triangular lattice. The spin-1/2
Heisenberg model on a triangular lattice has been of considerable 
recent interest, motivated largely in part by the discovery of the 
triangular lattice material Cs$_2$CuCl$_4$ and, in particular, 
to evidence that this material exhibits a two-dimensional 
fractionalized phase with deconfined spinons.\cite{coldea} 
However, besides being a toy model for understanding spin systems 
on a triangular lattice, this model is relevant to the 
quasi-one-dimensional material SrCu0$_2$.\cite{zigzagexp}

\begin{figure}[htb]
\scalebox{.285}{\includegraphics{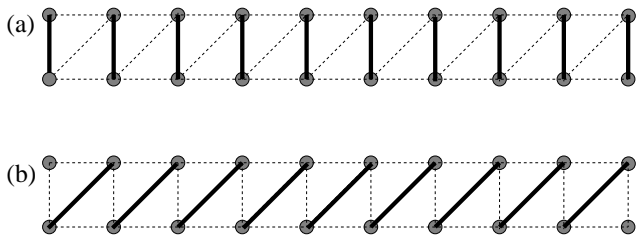}}
\caption{The two degenerate ground states of the zigzag ladder 
at the Majumdar-Ghosh point.}
\label{fig:zigzagGS}
\end{figure}

The properties of the zigzag ladder are well known when $J_{\bot}$ or $J_2$ 
vanishes, and also along the line $J_{\bot} = J_2$.
When $J_{\bot}$ or $J_2$ vanishes, the model reduces to the ordinary ladder.  
When $J_{\bot} = J_2$, the zigzag ladder is equivalent 
to the spin-1/2 frustrated Heisenberg chain with nearest-neighbor coupling
$J_{\bot}$ and next-nearest neighbor coupling $J_{\|}$. This chain model is 
known to be critical for $J_{\bot} > J_{\bot c}$ 
($J_{\bot c} = J_{\|}/0.241$),\cite{okamoto,eggert} being in the same 
universality class as the antiferromagnetic spin-1/2 Heisenberg chain.  
The spinons of the spin-1/2 chain acquire a gap (but they remain 
deconfined), and the ground state becomes doubly degenerate for 
$J_{\bot} = J_2 < J_{\bot c}$.\cite{haldane-j2} At the Majumdar-Ghosh (MG)
point\cite{majumdar} $J_{\bot} = J_2 = 2 J_{\|}$, the two degenerate ground 
states have a simple form in the thermodynamic limit, each consisting of 
decoupled singlets, as shown in Fig.~\ref{fig:zigzagGS}. It has been shown 
that these two ground states can be continuously related to the rung-singlet 
and Haldane phases.\cite{kolezhuk} For even weaker interchain couplings
$J_{\bot} = J_2 < 2 J_{\|}$, incommensurate oscillations appear in the 
short-range correlations.\cite{tonegawa,chitra,white-aff,aligia,legeza_incomm} 

%%%%%%%%%%%%%%%%%%%%

\subsubsection{Diagonal Ladder}

\begin{figure}[htb]
\scalebox{.31}{\includegraphics{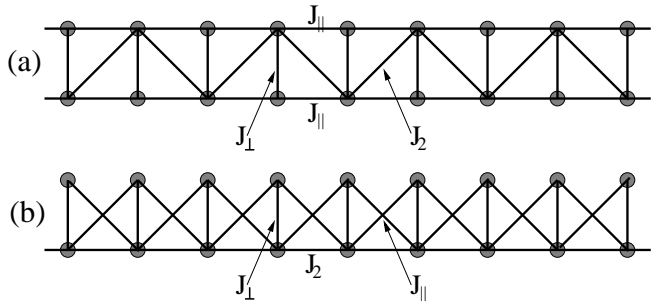}}
\caption{(a) The diagonal ladder model.
(b) Alternative representation of the diagonal ladder, obtained
by interchanging sites on every second rung.}
\label{fig:diagonal}
\end{figure}

Finally, we consider the model obtained when the couplings 
described both by \eqref{ordinaryladder} and \eqref{diagonal} are 
present simultaneously.  The model is shown in 
Fig.~\ref{fig:diagonal}(a) and is described by the Hamiltonian 
\begin{equation}
H_{\text{D}} = \sum_{i=1}^2H_0 + H_{\bot} + H_{\text{d}}  \,.
\label{eq:diagonal-2}
\end{equation}
For the rest of this work, we will refer to this ladder model as 
a {\sl diagonal ladder}.
Unlike the previous two ladder models which were invariant under 
translation by a single site, \eqref{eq:diagonal-2} is invariant 
under translation by two sites.  Furthermore, similar to the zigzag 
ladder, \eqref{eq:diagonal-2} is not invariant under the interchange 
of the legs.  However, it is invariant under the combined operation of 
leg interchange followed by a translation by a single site.  
Interestingly, this ladder model can, in fact, be transformed into
one which is invariant under translation by a single site, but 
lacking inversion symmetry.  This is accomplished by interchanging 
the sites on every second rung of the ladder, 
${\bf S}_{2l}^1 \leftrightarrow {\bf S}_{2l}^2$; the resulting 
model is shown in Fig.~\ref{fig:diagonal}. We are unaware of work 
which investigated the role of frustrating interactions in this model.

%%%%%%%%%%%%%%%%%%%%%%%%%%%%%%%%%%%%%%%%%
%%%%%%%%%%%%%%%%%%%%%%%%%%%%%%%%%%%%%%%%%

\section{Exact ground states}

An interesting feature of the models considered in this work
is that their exact ground states have a simple form for a
certain regime of parameters.  The key to establishing this
is decomposing the system into a set of ``triangles''.\cite{ss}
Here, we review the argument and discuss the ground states that arise.

Consider three $s=1/2$ spins arranged on a triangle. Suppose
the spins on sites 1 and 2 are coupled by an exchange coupling 
$J_1$; these two spins are coupled to the spin on site 3 by an
exchange coupling $J$. The Hamiltonian for the triangle is
\begin{equation}
H_{\rm triangle} = J_1 {\bf S}_1 \cdot {\bf S}_2 
+ J \big( {\bf S}_1 + {\bf S}_2\big) \cdot {\bf S}_3 \,.
\label{Htriangle}
\end{equation}
We are interested in the case where both $J$ and $J_1$ are 
antiferromagnetic; hence, the ground state is a doublet. There are 
two ways of achieving this: (a) a singlet across sites 1 and 2 with 
a free spin-1/2 on site 3, or (b) a triplet across sites 1 and 2 
added to the spin-1/2 on site 3 to form a spin-1/2. Let $[i,j]$ 
denote a singlet between sites $i$ and $j$.  Then, using that 
\begin{equation}
( {\bf S}_i + {\bf S}_j ) [i,j] = 0 \,,
\label{singlet}
\end{equation}
the energy of the state in (a) is found to be 
$E_{\text{a}} = - 3J_1/4$; the energy of state (b) is 
$E_{\text{b}} = - J + J_1/4$.  We see that the state in (a) 
--- the state with a singlet across sites 1 and 2 --- is the 
lowest-energy state if $J < J_1$. 

Using information from the previous paragraph, we now address
the ground states of the ladder models considered in this work.
We begin by considering the cross-coupled ladder model.  In
this model, four triangles can be assigned to each rung; the 
Hamiltonian can be written as
\begin{equation}
H_{\rm CC} = \sum_l \big( H_{l+}^{(1)} + H_{l-}^{(1)} 
+ H_{l+}^{(2)} + H_{l-}^{(2)}\big)  \,;
\label{Hcrosssumtriangles}
\end{equation}  
with
\begin{equation}   \begin{split}
H_{l+}^{(1)} &  = {\textstyle 
\frac{1}{4}}J_{\bot} {\bf S}_l^{(1)} \cdot{\bf S}_l^{(2)} 
+ {\textstyle \frac{1}{2}}J_{\|} {\bf S}_l^{(1)} \cdot{\bf S}_{l+1}^{(1)} 
+ {\textstyle \frac{1}{2}}J_{\text{X}} {\bf S}_l^{(2)} \cdot{\bf S}_{l+1}^{(1)}    
\,,   \\
H_{l-}^{(1)} &  = {\textstyle \frac{1}{4}}J_{\bot} {\bf S}_l^{(1)} \cdot{\bf S}_l^{(2)} 
+ {\textstyle \frac{1}{2}}J_{\|} {\bf S}_l^{(1)} \cdot{\bf S}_{l-1}^{(1)} 
+ {\textstyle \frac{1}{2}}J_{\text{X}} {\bf S}_l^{(2)} \cdot{\bf S}_{l-1}^{(1)}    
\,,   \\
H_{l+}^{(2)} &  = {\textstyle \frac{1}{4}}J_{\bot} {\bf S}_l^{(1)} \cdot{\bf S}_l^{(2)} 
+ {\textstyle \frac{1}{2}}J_{\|} {\bf S}_l^{(2)} \cdot{\bf S}_{l+1}^{(2)} 
+ {\textstyle \frac{1}{2}}J_{\text{X}} {\bf S}_l^{(1)} \cdot{\bf S}_{l+1}^{(2)}    
\,,   \\
H_{l-}^{(2)} &  = {\textstyle \frac{1}{4}}J_{\bot} {\bf S}_l^{(1)} \cdot{\bf S}_l^{(2)} 
+ {\textstyle \frac{1}{2}}J_{\|} {\bf S}_l^{(2)} \cdot{\bf S}_{l-1}^{(2)} 
+ {\textstyle \frac{1}{2}}J_{\text{X}} {\bf S}_l^{(1)} \cdot{\bf S}_{l-1}^{(2)}    
\,.  \end{split}    
\label{Hcrosstriangles}
\end{equation}
When $J_{\|} = J_{\text{X}}$, the terms in \eqref{Hcrosstriangles} can be 
written in the form of \eqref{Htriangle}. Then, using \eqref{singlet}, it 
follows that the state
\begin{equation}
| \psi \rangle = \prod_l [l1,l2] 
\label{simpleexact}
\end{equation}
where the spins on each rung form a singlet, is an exact eigenstate 
of the Hamiltonian in \eqref{Hcrosssumtriangles} with energy
\begin{equation}
E = - \frac{3}{4} J_{\bot} N  \,.
\label{simpleexactenergy}
\end{equation}
We would like to determine if and when \eqref{simpleexact} is 
the ground state. To do so, consider each of the triangles in 
\eqref{Hcrosstriangles} individually.  
If $J_{\bot}/2 > J_{\|}$, the lowest-energy state of each triangle 
(when considered individually) has a singlet between the spins on 
the same rung. It is known that if a Hamiltonian can be written as 
the sum of terms, the ground-state energy cannot be smaller than the 
sum of the lowest energies of its constituents. For the ground state 
of \eqref{Hcrosssumtriangles}, this gives the inequality
\begin{equation}
E_0 \geq - \frac{3}{4} J_{\bot} N  \,,
\label{bound}
\end{equation}
where $N$ is the number of rungs. Hence, we see that 
\eqref{simpleexactenergy} saturates the bound in \eqref{bound} when 
$J_{\bot}/2 > J_{\|}$; \eqref{simpleexact} is the ground state for 
$J_{\|} = J_{\text{X}}$ with $J_{\bot}/2 > J_{\|}$.\cite{martin,bosegayen}

We now consider the zigzag ladder.  It can be decomposed into two 
triangles per rung:
\begin{equation}
H_{\text{Z}} = \sum_l \big( H_{l+}^{(2)} + H_{l-}^{(1)}\big)  \,;
\end{equation}  
where
\begin{eqnarray}   
H_{l+}^{(2)} &  = & {\textstyle 
\frac{1}{2}}J_{\bot} {\bf S}_l^{(1)} \cdot{\bf S}_l^{(2)} 
+ J_{\|} {\bf S}_l^{(2)} \cdot{\bf S}_{l+1}^{(2)} 
+ {\textstyle \frac{1}{2}}J_2 {\bf S}_l^{(1)} \cdot{\bf S}_{l+1}^{(2)}    
\,,   \nonumber  \\[-1mm]
& &   \label{Hzztriangles}    \\[-1mm]
H_{l-}^{(1)} &  = & {\textstyle 
\frac{1}{2}}J_{\bot} {\bf S}_l^{(1)} \cdot{\bf S}_l^{(2)} 
+ J_{\|} {\bf S}_l^{(1)} \cdot{\bf S}_{l-1}^{(1)} 
+ {\textstyle \frac{1}{2}}J_2 {\bf S}_l^{(2)} \cdot{\bf S}_{l-1}^{(1)}    
\,.    \nonumber 
\end{eqnarray}
When $J_{\|} = J_2/2$, the terms in \eqref{Hzztriangles} can be 
written in the form of \eqref{Htriangle}; the energy of the state 
in \eqref{simpleexact} saturates the lower bound in \eqref{bound} 
for $J_{\bot}/2 > J_{\|}$.  Hence, \eqref{simpleexact} is the 
exact ground state for $J_{\|} = J_2/2$ with 
$J_{\bot}/2 > J_{\|}$.\cite{ss}
Note that another exact ground state of the zigzag ladder follows 
by symmetry. When $J_{\|} = J_{\bot}/2$ and $J_2/2 \geq J_{\|}$ 
the ground state is again a product of singlets, but this time 
they are formed diagonally between neighboring rungs:\cite{ss}
\begin{equation}
| \psi \rangle = \prod_l [l1,(l-1)2]  \,.
\label{simpleexact-2}  
\end{equation}

We now go on and consider the diagonal ladder.  The diagonal 
ladder can be decomposed into two triangles per rung, but the 
triangles have different orientations for even and odd rungs:
\begin{equation}
H_{\text{D}} = \sum_l \big( H_{(2l)+}^{(2)} + H_{(2l)-}^{(2)} + 
   H_{(2l+1)+}^{(1)} + H_{(2l+1)-}^{(1)} \big)  \,;
\end{equation}  
where now
\begin{equation}   \begin{split}
H_{(2l)+}^{(2)} &  = {\textstyle 
\frac{1}{2}}J_{\bot} {\bf S}_{2l}^{(1)} \cdot{\bf S}_{2l}^{(2)} 
+ J_{\|} {\bf S}_{2l}^{(2)} \cdot{\bf S}_{2l+1}^{(2)}  
\\  & \phantom{= }  + {\textstyle 
\frac{1}{2}}J_2 {\bf S}_{2l}^{(1)} \cdot{\bf S}_{2l+1}^{(2)}    
\,,    \\
H_{(2l)-}^{(2)} &  = {\textstyle 
\frac{1}{2}}J_{\bot} {\bf S}_{2l}^{(1)} \cdot{\bf S}_{2l}^{(2)} 
+ J_{\|} {\bf S}_{2l}^{(2)} \cdot{\bf S}_{2l-1}^{(2)}  \\
& \phantom{=} + {\textstyle 
\frac{1}{2}}J_2 {\bf S}_{2l}^{(1)} \cdot{\bf S}_{2l-1}^{(2)}    
\,,    \\
H_{(2l+1)+}^{(1)} &  = {\textstyle 
\frac{1}{2}}J_{\bot} {\bf S}_{2l+1}^{(1)} \cdot{\bf S}_{2l+1}^{(2)} 
+ J_{\|} {\bf S}_{2l+1}^{(1)} \cdot{\bf S}_{2l+2}^{(1)}  \\
& \phantom{=} + {\textstyle 
\frac{1}{2}}J_2 {\bf S}_{2l+1}^{(1)} \cdot{\bf S}_{2l+2}^{(2)}    
\,,  \\
H_{(2l+1)-}^{(1)} &  = {\textstyle 
\frac{1}{2}}J_{\bot} {\bf S}_{2l+1}^{(1)} \cdot{\bf S}_{2l+1}^{(2)} 
+ J_{\|} {\bf S}_{2l+1}^{(1)} \cdot{\bf S}_{2l}^{(1)}  \\
& \phantom{=} + {\textstyle 
\frac{1}{2}}J_2 {\bf S}_{2l+1}^{(2)} \cdot{\bf S}_{2l}^{(1)}    
\,.   \end{split} 
\label{Hdiagonaltriangles}   
\end{equation}
When $J_{\|} = J_2/2$, the terms in \eqref{Hdiagonaltriangles} can be 
written in the form of \eqref{Htriangle}; and thus \eqref{simpleexact} 
is the exact ground state for $J_{\|} = J_2/2$ with $J_{\bot}/2 \geq J_{\|}$.

%%%%%%%%%%%%%%%%%%%%%%%%%%%%%%%%%%%%%%%%%%
%%%%%%%%%%%%%%%%%%%%%%%%%%%%%%%%%%%%%%%%%%

\section{Topological order and spinons in ladder models}

>From the discussion in Secs.~II and III, the ladder models considered
in this work have similar ground-state properties in extended regions 
of parameter space --- their ground states are described by a collection 
of short-ranged valence bonds, separated by a gap to the excited states.  
However, the ground states of these ladder models, in fact, differ in 
a subtle way, namely in their {\sl topological order}.  More specifically, 
the number of valence bonds crossing an arbitrary vertical line is always
even in the rung-singlet phase, while the number is always odd in the 
Haldane phase.  This can be seen explicitly in the configurations shown
in Figs.~\ref{fig:ordinaryrvb}, \ref{fig:compositervb}, and \ref{fig:zigzagGS}.  
Hence a topological number $Q$ can be defined  by the parity of the 
number of short-range valence bonds crossing an arbitrary vertical 
line.\cite{gene} This $Q$, which is either even or odd, is a good quantum
number for short-range valence-bond states since the Hamiltonian has 
finite matrix elements only between configurations with the same $Q$.  
For long-range valence-bond states, however, the Hamiltonian mixes
the $Q={\text{even}}$ and $Q={\text{odd}}$ configurations; hence, no 
such topological distinction is possible.

It is worth noting that for open boundary conditions $Q = {\text{odd}}$ 
ground states have spin-1/2's localized at the ends of the ladder, while 
$Q = {\text{even}}$ states do not.  As can be seen from 
Figs.~\ref{fig:compositervb} and \ref{fig:zigzagGS}(b), these end spins 
occur for topological reasons.  They are analogous to the edge states in 
the quantum Hall effect; in general, the presence of such edge excitations 
is a signal of nontrivial topological order.\cite{wen}

It has also been pointed out\cite{gene} that the topological order of the 
valence bonds is related to the ``hidden order" present in two-leg Heisenberg 
spin ladders, namely {\sl string order}, analogous to the string order
in antiferromagnetic spin-1 chains.\cite{dennijs} This string order is 
detected by the two string order parameters\cite{gene,white-trans,nishiyama}
\begin{widetext}
\begin{eqnarray}
{\cal O}^{\alpha}_{\rm odd} & = & - \lim_{|i-j| \rightarrow \infty}
\left \langle (S^{\alpha}_{i,1}+S^{\alpha}_{i,2})~
\exp \left(i\pi~ \sum_{l=i+1}^{j-1}
 (S^{\alpha}_{l,1}+S^{\alpha}_{l,2}) \right)~
 (S^{\alpha}_{j,1}+S^{\alpha}_{j,2}) \right \rangle , \nonumber \\
{\cal O}^{\alpha}_{\rm even} & = & - \lim_{|i-j| \rightarrow \infty}
\left \langle (S^{\alpha}_{i+1,1}+S^{\alpha}_{i,2})~
\exp \left(i\pi~ \sum_{l=i+1}^{j-1}
 (S^{\alpha}_{l+1,1}+S^{\alpha}_{l,2}) \right)~
 (S^{\alpha}_{j+1,1}+S^{\alpha}_{j,2}) \right \rangle .
\label{stringorder}
\end{eqnarray}
\end{widetext}
It was shown\cite{shelton} that a slightly modified version of these string 
order parameters can be written in terms of the order and disorder fields 
of two Ising models --- ${\cal O}_{\rm odd}$ can be written in terms of the 
order fields $\sigma_1$ and $\sigma_2$; ${\cal O}_{\rm even}$ can be 
written in terms of the disorder fields $\mu_1$ and $\mu_2$. 
Hence, ${\cal O}_{\rm odd}$ and ${\cal O}_{\rm even}$ cannot be nonzero 
simultaneously, as they are dual to each other.
It was observed that models with $Q = {\text{odd}}$ ground states have 
${\cal O}_{\rm odd} \neq 0$, while 
models with $Q = {\text{even}}$ ground states were observed to have
${\cal O}_{\rm even} \neq 0$.\cite{gene,fath} 
Hence, the string order parameters detect the topological order 
of the valence bonds.

\begin{figure}[ht]
\scalebox{.25}{\includegraphics{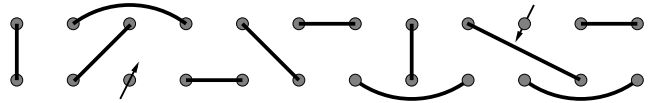}}
\caption{Spinons separating topologically distinct regions.}
\label{fig:spinon}
\end{figure}

This topological order has important consequences for the excitation 
spectrum.  In these models with short-range valence-bond ground states,
the simplest excitation is generated by breaking one of the valence 
bonds, promoting it to a triplet. 
An interesting and important question is how this excited state 
propagates.  More specifically, does the triplet propagate coherently, 
or does it ``break apart" so that the individual spins forming the 
triplet --- referred to as {\sl spinons} --- propagate independently?
In both the rung-singlet and Haldane phases, it is known that the 
triplet propagates coherently.  The reason for this can be understood 
looking at Fig.~\ref{fig:spinon} and counting the number of valence bonds 
crossing a vertical line. When a valence bond is broken in a state with 
$Q={\text{even}}$, if this triplet breaks apart it leaves a string of valence 
bonds with the ``wrong" topology in the intermediate region.  This gives 
rise to an increase in the local energy, which is proportional to the 
distance between the two spinons; as a result, the spinons are confined 
into a (gapped) spin-1 magnon. 
(A similar situation occurs in the $Q={\text{odd}}$ Haldane phase.) 

>From Fig.~\ref{fig:spinon}, one also sees that if the two topologically 
distinct ground states can be made degenerate, one can expect spinons to 
be deconfined.  This is because the string of ``wrong" valence bonds 
between the two spinons would not give an increase in energy. Thus, the 
degeneracy provides the "liquidity" necessary for spinon 
deconfinement.\cite{liquidity}
Indeed, as discussed in Sec.~II, this is known to happen in the zigzag 
ladder due to the frustrating interaction. 
It is reasonable to expect similar phenomena could occur generically 
--- frustrating interactions could tune the topologically distinct 
rung-singlet and Haldane ground states to be degenerate, so that spinons 
are deconfined and propagate as elementary excitations.
While this expectation is reasonable, as will be discussed in detail
below, there are other possible phases --- namely dimerized phases ---
which could intervene.  Moreover, we will see that the system's
geometry plays a crucial role in determining whether the system may
dimerize or not.

%%%%%%%%%%%%%%%%%%%%%%%%%%%%%%%%%%%%%%%%%%
%%%%%%%%%%%%%%%%%%%%%%%%%%%%%%%%%%%%%%%%%%

\section{Weak-Coupling analysis}

It is useful to consider the physics in the limit where the interchain 
coupling is weak, as controlled analytic calculations are possible. 
More specifically, we start with two decoupled spin-1/2 chains 
and consider the interchain coupling as a perturbation.
Provided that there is no phase transition, the low-energy
Hamiltonian deduced in the weak-coupling limit is valid even
at strong coupling, albeit with renormalized parameters.
This approach has been utilized in various other works;
besides providing a detailed understanding of the properties
of the two-leg ladder and its relation to the spin-1 
chain,\cite{shelton} it has been effective at uncovering 
and illucidating the phenomena that can arise in these 
systems.\cite{nt,nge,orignac,plateau,allen,oleg}

We begin this section by briefly describing the formalism, mainly 
to establish our conventions.  (More detailed accounts can be found, 
e.g., in Ref.~\onlinecite{cftbook}.)  We go on to deduce the effective 
low-energy Hamiltonians for the models considered in this work, 
and then discuss the physics contained in these effective 
Hamiltonians. In particular, we discuss the role of the various irrelevant
operators that arise.  These operators were ignored in most 
previous works, but were recently argued to affect the physics 
qualitatively in some situations.\cite{oleg} As will be elaborated on below, 
quantum fluctuations can give rise to several different behaviors.
To understand the physics of these quantum fluctuations, we use 
the renormalization group (RG) and examine the behavior of the
system under a change of scale.

%%%%%%%%%%%%%%%%%%%%%%%%%%%%%%%%%%%%%%%%%%

\subsection{Formalism}

The low-energy properties of the spin-1/2 chain are described
by an SU(2)$_1$ WZW model with Hamiltonian
\begin{equation}
H = \frac{v}{2\pi} \int dx \left( {\bf J}_R \cdot {\bf J}_R 
+ {\bf J}_L \cdot {\bf J}_L \right) ,
\label{WZWhamiltonian}  
\end{equation}
where the velocity $v$ is related to $J_{\|}$, and
${\bf J}_R$ and ${\bf J}_L$ are currents satisfying the SU(2)$_1$ 
Kac-Moody operator product expansion (OPE)
\begin{eqnarray}
:J^{\alpha}_R(z):: J^{\beta}_R(w): &=&
\frac{\delta_{\alpha,\beta}}{2 (z-w)^2} +
\frac{\text{i} \epsilon^{\alpha \beta \gamma}}{2(z-w)} J^{\gamma}_R(w) \, ,
\nonumber  \\[-1mm]
   & & \label{kacmoodyleft}   \\[-1mm]
:J^{\alpha}_L(\overline{z}):: J^{\beta}_L(\overline{w}): &=& 
\frac{\delta_{\alpha,\beta}}{2 (\overline{z} - \overline{w})^2} 
+ \frac{\text{i} \epsilon^{\alpha \beta \gamma}}
{2(\overline{z} - \overline{w})} J^{\gamma}_L(w) \, ,
\nonumber
\end{eqnarray}
with $z = v\tau + {\text{i}}x$ and ${\overline z} = v\tau - {\text{i}}x$.
The SU(2)$_1$ WZW model has a single primary (matrix)
field $g(z,{\overline z})$ of dimension $(1/4,1/4)$.  
($g$ is the field appearing in the $\sigma$-model representation.) 
Physical operators of the spin-1/2 chain are given by 
combinations of the components $g$ --- the 
{\it staggered magnetization} $n^{\alpha}(z,{\overline z})$ 
and the {\it dimerization} $\epsilon(z,{\overline z})$: 
\begin{equation}
n^{\alpha}(z,{\overline z}) 
= {\rm Tr}[\sigma^{\alpha} g(z,{\overline z})]
 \,, \ \ \ 
\epsilon(z,{\overline z}) 
= {\rm Tr}[g(z,{\overline z})]  \,,
\end{equation}
where the $\{\sigma^{\alpha}\}$ are Pauli matrices.
The $n^{\alpha}$ and $\epsilon$ fields have leading
short-distance behavior
\begin{eqnarray}
:n^{\alpha}(z)::n^{\beta}(w): & = & 
\frac{\delta_{\alpha \beta}}{|z-w|}  \,,   
\nonumber \\
:\epsilon(z)::\epsilon(w): & = & \frac{1}{|z-w|}  \,,  
\label{primaryOPE}  \\ 
:n^{\alpha}(z)::\epsilon(w): & = & 0  \,.  \nonumber
\end{eqnarray}
Furthermore, being linear combinations of the components of $g$, 
their OPE's with the currents are
\begin{eqnarray}
:J^{\alpha}_R(z):: n^{\beta}(w):
& = & \frac{\text{i}}{2(z-w)} \left[
\epsilon^{\alpha \beta \gamma} n^{\gamma}(w)
- \delta_{\alpha \beta} \epsilon(w) \right]  ,
\nonumber  \\
:J^{\alpha}_L(z):: n^{\beta}(w):
& = & \frac{\text{i}}{2(\overline{z} - \overline{w})} \left[ 
\epsilon^{\alpha \beta \gamma} n^{\gamma}(w)
+ \delta_{\alpha \beta} \epsilon(w) \right]  ,
\nonumber \\
:J^{\alpha}_R(z):: \epsilon(w):
& = & \frac{\text{i}}{2(z-w)} n^{\alpha}(w)  \,,  
\label{Jprimary}  \\
:J^{\alpha}_L(z):: \epsilon(w):
& = & \frac{-{\text{i}}}{2(\overline{z} - \overline{w})} 
n^{\alpha}(w)  \,.  \nonumber
\end{eqnarray}
[In Eqs.~\eqref{primaryOPE} and \eqref{Jprimary},
$n^{\alpha}(z) \equiv n^{\alpha}(z,{\overline z})$ and 
$\epsilon(z) \equiv \epsilon(z,{\overline z})$.]

At low energies, the spin operator on chain $i$, 
${\bf S}^{(i)}_l$, can be written as
\begin{equation}
\frac{{\bf S}^{(i)}_l}{a} = \frac{1}{2\pi} 
\left( {\bf J}_{iR}(x) + {\bf J}_{iL}(x) \right) 
+ (-1)^l \frac{M}{2\pi a^{1/2}} {\bf n}_i(x)  \, ,
\label{uniplusstagg}
\end{equation}
where $a$ is an ultraviolet regulator ($a \sim$ the lattice spacing), 
and $M$ is a nonuniversal ${\cal O}(1)$  constant.
In what follows, we will need to know the behavior of the
fields upon translation by a single site.  
>From the invariance of the spin operator ${\bf S}^{(i)}_l$
under translation we deduce how ${\bf J}_{iR}(x)$,
${\bf J}_{iL}(x)$, and ${\bf n}_i(x)$ transform.  Then,
as all of the terms on the right-hand-side of \eqref{Jprimary} 
must transform in the same way upon translation, we deduce 
how $\epsilon_i(x)$ transforms.  Hence, we arrive at
\begin{eqnarray}
{\bf J}_{iR/L}(x) & \rightarrow & {\bf J}_{iR/L}(x)  
\,,  \\
{\bf n}_i(x) \rightarrow -{\bf n}_i(x)  \,, &   & \
\epsilon_i(x) \rightarrow -\epsilon_i(x) \,.
\nonumber
\end{eqnarray}

%%%%%%%%%%%%%%%%%%%%%%%%%%%%%%%%%%%%%%%%%%

\subsection{Operators and Phases: Topological Order,
Dimer Order, and Deconfined Spinons}

As described above, starting from two spin-1/2 chains, we 
are interested in the fate of the system upon turning on an interchain 
coupling. In this subsection, we discuss the various operators that
can arise in the low-energy Hamiltonian(s); we discuss the 
physics that these operators give rise to.

The low-energy Hamiltonians are dictated by symmetry.
The models we are considering are all SU(2) symmetric;
hence, all operators appearing must transform as SU(2) 
scalars.  Therefore, if only relevant and marginal operators 
are considered, the operators
\begin{eqnarray}
& & \epsilon_1 ~ {\rm and} ~ \epsilon_2 \,, \ \   
\epsilon_1 \epsilon_2 \,, \ \ 
{\bf n}_1 \cdot {\bf n}_2   \,, \ \ 
{\bf n}_1 \partial_x {\bf n}_2 \,, \ \ 
\epsilon_1 \partial_x \epsilon_2 \,,
\nonumber  \\
& & \left( {\bf J}_{1R} 
+ {\bf J}_{1L} \right) \cdot {\bf n}_2 \,, \quad
 \left( {\bf J}_{2R} + {\bf J}_{2L} \right) 
 \cdot {\bf n}_1  \,, 
\label{potential}  \\
& & ({\bf J}_{1R} \cdot {\bf J}_{2L} 
+ {\bf J}_{2R} \cdot {\bf J}_{1L}) \,, \ \ 
({\bf J}_{1R} \cdot {\bf J}_{1L} 
+ {\bf J}_{2L} \cdot {\bf J}_{2R}) 
\nonumber
\end{eqnarray}
have the potential of appearing in the low-energy
Hamiltonians. Furthermore, we will not consider the terms 
${\bf J}_{1R} \cdot {\bf J}_{2R}$ and 
${\bf J}_{1L} \cdot {\bf J}_{2L}$.  
Even though these terms have dimension two (and, hence, 
are marginal), they couple excitations moving in the same 
direction on both chains and give rise only to small 
quantitative corrections. As discussed below, other 
symmetries of the various models will further restrict the 
operators allowed in their low-energy Hamiltonians.

The physics is determined by the operator which flows to strong
coupling first under the RG. When the ${\bf n}_1 \cdot {\bf n}_2$ 
term determines the physics, the resulting phases have gapped 
magnons with spinons being confined.\cite{shelton}
[The ${\bf n}_1 \cdot {\bf n}_2$ term, in fact, gives rise to a
strong confining potential which binds together spinons from the 
two chains to form a gapped magnon.]  Furthermore, these phases have 
topological order; information about this topological order is 
contained in the sign of the ${\bf n}_1 \cdot {\bf n}_2$ term's 
coefficient.
Another possibility is if the physics is determined by the
(${\bf J}_{1R} \cdot {\bf J}_{2L} + {\bf J}_{2R} \cdot {\bf J}_{1L}$)
term.  When this occurs, it has been shown that the resulting
phase is a fractionalized phase with deconfined spinons.\cite{allen}
Finally, if the physics is determined by the 
$\epsilon_1$ and $\epsilon_2$ or $\epsilon_1 \epsilon_2$ 
term(s), the result is a spontaneously dimerized phase.  

There are other operators in Eq.~\eqref{potential}.  However, these 
operators do not determine the phases that arise, though they modify 
the properties within a particular phase.  In particular, 
$({\bf J}_{1R} \cdot {\bf J}_{1L} + {\bf J}_{2L} \cdot {\bf J}_{2R})$
is the marginally irrelevant operator present in the two spin-1/2 
chains.\cite{leshouches}
Furthermore, ${\bf n}_1 \cdot \partial_x {\bf n}_2$  and 
$\epsilon_1 \partial_x \epsilon_2$ are believed to give rise to 
incommensurate correlations.\cite{allen,plateau}  
Finally, as will be seen below, the 
$( {\bf J}_{1R} + {\bf J}_{1L}) \cdot {\bf n}_2$ and 
$( {\bf J}_{2R} + {\bf J}_{2L}) \cdot {\bf n}_1$ are always 
subleading to ${\bf n}_1 \cdot {\bf n}_2$ and $\epsilon_1 \epsilon_2$ 
and, hence, do not determine the phases that arise.

%%%%%%%%%%%%%%%%%%%%%%%%%%%%%%%%%%%%%%%%%%%%

\subsection{Effective Low-Energy Hamiltonians}

Here we deduce the effective low-energy Hamiltonians for the 
frustrated ladder models considered in this work. We then derive 
renormalization group (RG) equations describing how the parameters 
in the low-energy Hamiltonian evolve under a change of scale.
Finally, we derive the initial values of the parameters from 
the microscopic models in the limit of weak interchain coupling.

%%%%%%%%%%%%%%%%%%%%%%%%%%%%%%%%%%%%%%%%%%%%%

\subsubsection{Cross-Coupled Ladder}

Besides being SU(2)-symmetric, the cross-coupled ladder is 
invariant under translation by one site and also inversion 
about the center of the ladder: leg 1 $\leftrightarrow$ leg 2.
Translation by one site forbids $\epsilon_1$ and $\epsilon_2$ 
from appearing alone; it also forbids the operators
$( {\bf J}_{1R} + {\bf J}_{1L}) \cdot {\bf n}_2$  
and 
$( {\bf J}_{2R} + {\bf J}_{2L}) \cdot {\bf n}_1$.
Inversion about the center forces operators from 
leg 1 and leg 2 to appear symmetrically; in particular, it
forbids the operators 
${\bf n}_1 \partial_x {\bf n}_2$ and $\epsilon_1 \partial_x \epsilon_2$
from appearing.
These symmetries constrain the effective low-energy Hamiltonian 
to have the form
\begin{eqnarray}
& & H = \int dx \Big[  \gamma~ ({\bf J}_{1R} \cdot {\bf J}_{1L} 
+ {\bf J}_{2L} \cdot {\bf J}_{2R})     
\label{effectivecross}  \\
& & + \lambda~ ({\bf J}_{1R} \cdot {\bf J}_{2L} 
+ {\bf J}_{1L} \cdot {\bf J}_{2R})  
+ \frac{g}{a}~ {\bf n}_1 \cdot {\bf n}_2 
+ \frac{\beta}{a}~ \epsilon_1 \epsilon_2  \Big] \,.
\nonumber
\end{eqnarray}

To understand the physics contained in \eqref{effectivecross}, 
we use the RG and investigate the behavior under a transformation of scale.
Using the OPE's in \eqref{kacmoodyleft}, \eqref{primaryOPE}, 
and \eqref{Jprimary}, we deduce the RG equations for the parameters
to be\cite{cardybook} 
\begin{eqnarray}
\frac{d \gamma}{dl} & = & \gamma^2 - \frac{1}{2} g^2 
+ \frac{1}{2} \beta^2 \,,  \nonumber \\
\frac{d \lambda}{dl} & = & \lambda^2 + g^2 - g \beta  \,,
\\
\frac{d g}{dl} & = & g - \frac{1}{2} \gamma g
+ \lambda g - \frac{1}{2} \lambda \beta \,,
\nonumber \\
\frac{d \beta}{dl} & = & \beta + \frac{3}{2} \gamma \beta
- \frac{3}{2} \lambda g  \,.
\nonumber
\end{eqnarray}
>From the structure of the RG equations, we see that if either
$g$ or $\beta$ is nonzero, both ${\bf n}_1 \cdot {\bf n}_2$ 
and $\epsilon_1 \epsilon_2$ will be generated upon renormalization.
However, if both $g =0$ and $\beta=0$, these operators will not appear.
Hence, one would expect the relevant ${\bf n}_1 \cdot {\bf n}_2$ 
or $\epsilon_1 \epsilon_2$ terms to determine the physics under most 
situations. If both of these terms are suppressed, the physics would be
determined by the marginally relevant 
$({\bf J}_{1R} \cdot {\bf J}_{2L} + {\bf J}_{1L} \cdot {\bf J}_{2R})$.

To deduce the values of the parameters in \eqref{effectivecross}, 
we insert \eqref{uniplusstagg} into \eqref{ordinaryladder} 
and \eqref{J1J2}; we obtain 
\begin{eqnarray}
H_{\bot} + H_{\text{X}} & = & \int \frac{dx}{(2\pi)^2} \Big[
(J_{\bot} - 2 J_{\text{X}}) \frac{M^2}{a} {\bf n}_1 \cdot {\bf n}_2 
\nonumber \\
& - & 2J_{\text{X}} \frac{M^2}{a} \sum_{m=1}^{\infty} \frac{a^{2m}}{(2m)!}
  {\bf n}_1 \partial_x^{2m} {\bf n}_2 
\label{lowenergycross}  \\
& + & (J_{\bot} + 2 J_{\text{X}})  ({\bf J}_{1R} \cdot {\bf J}_{2L} 
+ {\bf J}_{1L} \cdot {\bf J}_{2R}) \Big] \,  .  \nonumber  
% \label{lowenergycross}  \\
\end{eqnarray}
A lattice Hamiltonian gives rise to a continuum field theory 
with an infinite number of operators.\cite{leshouches}
Here, besides ${\bf n}_1 \cdot {\bf n}_2$ and  
$({\bf J}_{1R} \cdot {\bf J}_{2L} + {\bf J}_{1L} \cdot {\bf J}_{2R})$,
an infinite number of irrelevant operators of the form 
${\bf n}_1 \cdot \partial_x^{2m} {\bf n}_2$ appear.
>From Eq.~\eqref{lowenergycross}, the relevant operator 
${\bf n}_1 \cdot {\bf n}_2$ determines the physics for generic 
values of $J_{||}$ and $J_{\bot}$. When the 
${\bf n}_1 \cdot {\bf n}_2$ term is suppressed, it
appears the physics is determined by the 
(${\bf J}_{1R} \cdot {\bf J}_{2L} + {\bf J}_{1L} \cdot {\bf J}_{2R}$) 
term. However, things are more subtle --- the irrelevant 
${\bf n}_1 \cdot \partial_x^{2m} {\bf n}_2$ can generate
the relevant ${\bf n}_1 \cdot {\bf n}_2$ and $\epsilon_1 \epsilon_2$
when integrated out.\cite{oleg}
Indeed, focussing on the leading irrelevant operator
${\bf n}_1 \cdot \partial_x^{2} {\bf n}_2$, the term
\begin{eqnarray}  
 (J_{\bot}&+&2J_{\text{X}})  \int \frac{d^2x_1}{(2\pi)^2} 
({\bf J}_{1R} \cdot {\bf J}_{2L} 
+ {\bf J}_{1L} \cdot {\bf J}_{2R})  \nonumber   \\ 
& & \times   J_{\text{X}} M^2 a \int \frac{d^2x_2}{(2\pi)^2}  
{\bf n}_1 \cdot \partial_x^{2} {\bf n}_2 
\end{eqnarray}
appears in the partition function at second order.  
Using the OPE's in \eqref{Jprimary}, 
\begin{equation}
\frac{J_{\text{X}} (J_{\bot} + 2J_{\text{X}}) M^2}{(2\pi)^2 a}~ 
\left[ \frac{1}{2} {\bf n}_1 \cdot {\bf n}_2
- \frac{3}{4} \epsilon_1 \epsilon_2 \right]
\end{equation}
is generated in the low-energy Hamiltonian.  Hence, to leading 
order we deduce 
\begin{eqnarray}
& & \lambda = \frac{J_{\bot} + 2J_{\text{X}}}{(2\pi)^2}  \,, \ \ \ 
\beta = -\frac{3}{8\pi}\frac{(J_{\bot} + 2J_{\text{X}})J_{\text{X}} M^2}{(2\pi)^2} 
\,, \nonumber \\
& & g = \frac{(J_{\bot} - 2J_{\text{X}}) M^2}{(2\pi)^2} 
+ \frac{J_{\text{X}} M^2}{4\pi} \frac{J_{\bot} + 2J_{\text{X}}}{(2\pi)^2} \,.
\label{initialcross}
\end{eqnarray}

As mentioned above, the relevant operator 
${\bf n}_1 \cdot {\bf n}_2$ determines the physics for generic 
values of $J_{||}$ and $J_{\bot}$. 
Now when $g \rightarrow 0$ in \eqref{initialcross}, the relevant 
$\epsilon_1 \epsilon_2$ is present (from integrating out the 
irrelevant operators); one may expect it to determine the physics.  
However, its coefficient is significantly smaller than the 
(${\bf J}_{1R} \cdot {\bf J}_{2L} + {\bf J}_{1L} \cdot {\bf J}_{2R}$) 
term.  Therefore, a subtle competition between the two interactions 
should be expected.  This competition will be investigated in detail in our 
numerical calculations.

%%%%%%%%%%%%%%%%%%%%%%%%%%%%%%%%%%%%%%%%%%%%

\subsubsection{Zigzag Ladder}

Like the cross-coupled ladder, the zigzag ladder is invariant
under translation by a single site.  However, it lacks the 
symmetry of inversion about the center of the ladder.  Hence, 
the operators from leg 1 and leg 2 do not have to appear 
symmetrically --- now the operators
${\bf n}_1 \partial_x {\bf n}_2$ and $\epsilon_1 \partial_x \epsilon_2$
are allowed. The effective low-energy Hamiltonian has the form
\begin{eqnarray}
&  &  H  = \int dx \Big[  \gamma~ ({\bf J}_{1R} \cdot {\bf J}_{1L} 
+ {\bf J}_{2L} \cdot {\bf J}_{2R})   \nonumber  \\
& & \hspace{.53in} +~ \lambda~ ({\bf J}_{1R} \cdot {\bf J}_{2L} 
+ {\bf J}_{1L} \cdot {\bf J}_{2R})   
\label{effectivezigzag}   \\
& & + \frac{g}{a}~ {\bf n}_1 \cdot {\bf n}_2 
+ \frac{\beta}{a}~ \epsilon_1 \epsilon_2 
+ g_1~ {\bf n}_1 \cdot \partial_x {\bf n}_2 
+ \beta_1~ \epsilon_1 \partial_x \epsilon_2  \Big] \,.  \nonumber    
\end{eqnarray}

As with the cross-coupled ladder, we investigate the physics of
Eq.~\eqref{effectivezigzag} using the RG.  Using the OPE's, we deduce 
the RG equations for the parameters to be
\begin{eqnarray}
\frac{d \gamma}{dl} & = & \gamma^2 - \frac{1}{2} g^2 
+ \frac{1}{2} \beta^2  + \frac{1}{4} g_1^2 - \frac{1}{4} \beta_1^2 \,,
\nonumber \\
\frac{d \lambda}{dl} & = & \lambda^2 + g^2 - g \beta
+ \frac{1}{2} g_1^2 - \frac{1}{2} g_1 \beta_1  \,,
\nonumber  \\
\frac{d g}{dl} & = & g - \frac{1}{2} \gamma g 
+ \lambda g - \frac{1}{2} \lambda \beta  \,,
\label{zigzagRG}  \\
\frac{d \beta}{dl} & = & \beta + \frac{3}{2} \gamma \beta 
- \frac{3}{2} \lambda g   \,,
\nonumber \\
\frac{d g_1}{dl} & = & - \frac{1}{2} \gamma g_1 
+ \lambda g_1 - \frac{1}{2} \lambda \beta_1  \,,
\nonumber \\
\frac{d \beta_1}{dl} & = & \frac{3}{2} \gamma \beta_1
- \frac{3}{2} \lambda g_1   \,.   \nonumber
\end{eqnarray}
As with the cross-coupled ladder, 
${\bf n}_1 \cdot {\bf n}_2$ and $\epsilon_1 \epsilon_2$ will be 
generated upon renormalization unless both $g =0$ and $\beta =0$.
Similarly, ${\bf n}_1 \cdot \partial {\bf n}_2$ and 
$\epsilon_1 \partial \epsilon_2$ will appear unless both 
$g_1 = 0$ and $\beta_1 =0$.
Hence, as with the cross-coupled ladder, one would expect the 
${\bf n}_1 \cdot {\bf n}_2$ or $\epsilon_1 \epsilon_2$ terms 
to determine the physics under most situations;
if they are suppressed, the physics would be determined by 
$({\bf J}_{1R} \cdot {\bf J}_{2L} + {\bf J}_{1L} \cdot {\bf J}_{2R})$.

To deduce the values of the parameters in Eq.~\eqref{effectivezigzag}, 
we insert Eq.~\eqref{uniplusstagg} into Eqs.~\eqref{ordinaryladder} and 
\eqref{triangular}; we obtain
\begin{eqnarray}
& & H_{\bot} + H_{\text{z}} = \int \frac{dx}{(2\pi)^2} \Big\{
(J_{\bot} - J_2) \frac{M^2}{a} {\bf n}_1 \cdot {\bf n}_2 
\nonumber \\
& & \ \ \ +~ \frac{M^2}{a} \sum_{m=1}^{\infty} 
\frac{1}{m!} \left(\frac{a}{2}\right)^{m} \left[ J_{\bot} (-1)^m - J_2 \right] 
{\bf n}_1 \partial_x^{m} {\bf n}_2
\nonumber \\
& & \ \ \ +~ (J_{\bot} + J_2)~ ({\bf J}_{1R} \cdot {\bf J}_{1L} 
+ {\bf J}_{2L} \cdot {\bf J}_{2R}) \Big\}  \,.    
\label{lowenergyzigzag} 
\end{eqnarray}
As we saw with the cross-coupled ladder, the lattice Hamiltonian
gives rise to an infinite number of operators in the field theory.
Similar to the cross-coupled ladder, the relevant  
${\bf n}_1 \cdot {\bf n}_2$ determines the physics for generic 
values of $J_{||}$ and $J_{\bot}$.
Furthermore, when the ${\bf n}_1 \cdot {\bf n}_2$ term is 
suppressed, the irrelevant operators may generate terms which 
determine the physics.
Interestingly, for this model all the terms of the form 
${\bf n}_1 \cdot \partial_x^{2m} {\bf n}_2$ have their
coefficient proportional to ($J_{\bot} - J_2$).  Hence, all the 
operators of the form ${\bf n}_1 \cdot \partial_x^{2m} {\bf n}_2$ 
are fine-tuned away along the line $J_{\bot} = J_2$; only terms of 
the form ${\bf n}_1 \cdot \partial_x^{2m+1} {\bf n}_2$ are
present. As there are only derivatives of odd power, they are unable 
to generate terms like ${\bf n}_1 \cdot {\bf n}_2$ or
$\epsilon_1 \epsilon_2$.
Therefore, for this model the irrelevant operators do not change 
the physics (though they give rise to small quantitative changes, e.g., 
in the size of energy gaps);  to leading order we deduce the parameters 
in  Eq.~\eqref{effectivezigzag} to be
\begin{eqnarray}
& & \lambda = \frac{J_{\bot} + J_2}{(2\pi)^2}  \,, \ \ \ 
g = \frac{(J_{\bot} - J_2) M^2}{(2\pi)^2}  \,, \ \ \ 
\beta = 0 \,,    \nonumber   \\  
& &  \ \ \ \ \ \ \ \  g_1 = - \frac{(J_{\bot} + J_2) M^2}{2 (2\pi)^2} 
\,, \ \  \  \beta_1 = 0  \,.
\label{initialzigzag}
\end{eqnarray}

%%%%%%%%%%%%%%%%%%%%%%%%%%%%%%%%%%%%%%

\subsubsection{Diagonal Ladder}

Like the previous two models, the diagonal ladder has translation 
invariance. However, unlike the cross-coupled and zigzag ladders which 
are invariant under translation by a single site, the diagonal ladder 
is invariant under translation by two sites.  Furthermore, while 
this model is not invariant under inversion about the center of 
the ladder (leg 1 $\leftrightarrow$ leg 2), it is invariant under 
inversion compounded by translation by one site.
Due to the invariance under translation by two sites, the
operators $\epsilon_1$, $\epsilon_2$, 
$( {\bf J}_{1R} + {\bf J}_{1L}) \cdot {\bf n}_2$  
and 
$( {\bf J}_{2R} + {\bf J}_{2L}) \cdot {\bf n}_1$ are now allowed.  
However, the symmetry of inversion compounded by translation by one 
site constrains them to appear in the combination 
($\epsilon_1 - \epsilon_2$) and 
[$( {\bf J}_{1R} + {\bf J}_{1L}) \cdot {\bf n}_2 - ( 
{\bf J}_{2R} + {\bf J}_{2L}) \cdot {\bf n}_1$].
Hence, the effective low-energy Hamiltonian has the form
\begin{eqnarray}
& & H = \int dx \Big[ 
\frac{\alpha}{a^{3/2}} \left( \epsilon_1 - \epsilon_2 \right) 
+ \gamma~ ({\bf J}_{1R} \cdot {\bf J}_{1L} 
+ {\bf J}_{2L} \cdot {\bf J}_{2R})     
\nonumber \\
& & + \lambda~ ({\bf J}_{1R} \cdot {\bf J}_{2L} 
+ {\bf J}_{1L} \cdot {\bf J}_{2R})  
+ \frac{g}{a}~ {\bf n}_1 \cdot {\bf n}_2 
+ \frac{\beta}{a}~ \epsilon_1 \epsilon_2 
\nonumber \\
& & + \frac{\kappa}{a^{1/2}} \left[ \left( {\bf J}_{1R} 
+ {\bf J}_{1L} \right) \cdot {\bf n}_2 
- \left( {\bf J}_{2R} + {\bf J}_{2L} \right) 
\cdot {\bf n}_1  \right]  \Big] \,  .  
\label{effectivediagonal}   
\end{eqnarray}

As before, we use the OPE's to deduce the RG equations for 
the parameters; we find
\begin{eqnarray}
\frac{d \gamma}{dl} & = & \gamma^2 - \frac{1}{2} g^2 
+ \frac{1}{2} \beta^2  + \frac{1}{2} \alpha^2 \,,
\nonumber \\
\frac{d \lambda}{dl} & = & \lambda^2 + g^2 - g \beta \,,
\nonumber  \\
\frac{d g}{dl} & = & g - \frac{1}{2} \gamma g 
+ \lambda g - \frac{1}{2} \lambda \beta - \kappa^2 \,,
\label{diagonalRG}  \\
\frac{d \beta}{dl} & = & \beta + \frac{3}{2} \gamma \beta
- \frac{3}{2} \lambda g + \alpha^2 - \frac{3}{2} \kappa^2  \,,
\nonumber \\
\frac{d \alpha}{dl} & = & \frac{3}{2} \alpha 
+ \frac{3}{4} \gamma \alpha + \beta \alpha  \,,
\nonumber \\
\frac{d \kappa}{dl} & = & \frac{1}{2} \kappa
- \frac{1}{4} \gamma \kappa + \frac{1}{4} \lambda \kappa
- g \kappa + \frac{1}{2} \beta \kappa  \,.    \nonumber
\end{eqnarray}
For this ladder model, the operator ($\epsilon_1 - \epsilon_2$) 
is allowed; if its initial coefficient is zero, this term will
never be generated.
Furthermore, we must have $\alpha =0$, $g =0$, $\beta =0$, 
and $\kappa =0$ in order for all the relevant operators to
be banished.  In particular, if any of these are nonzero,
the ${\bf n}_1 \cdot {\bf n}_2$ and $\epsilon_1 \epsilon_2$ 
terms will be generated.

To deduce the values of the parameters in Eq.~\eqref{effectivediagonal}, 
we insert \eqref{uniplusstagg} into \eqref{eq:diagonal-2}; 
we obtain
\begin{eqnarray}
& & H_{\bot} + H_{\text{d}} = \int \frac{dx}{(2\pi)^2} \Big\{
(J_{\bot} - J_2) \frac{M^2}{a} {\bf n}_1 \cdot {\bf n}_2 
\nonumber \\
& & \ \ \ -~ J_2 \frac{M^2}{a} \sum_{m=1}^{\infty} \frac{a^{2m}}{(2m)!}
  {\bf n}_1 \cdot \partial_x^{2m} {\bf n}_2 
\label{lowenergydiagonal}  \\
& & \ \ \ +~ (J_{\bot} + J_2) ~ ({\bf J}_{1R} \cdot {\bf J}_{1L} 
+ {\bf J}_{2L} \cdot {\bf J}_{2R})   
\nonumber \\
& & \ \ \ +~ \frac{J_2 M}{a^{1/2}} \left[ \left( {\bf J}_{1R} 
+ {\bf J}_{1L} \right) \cdot {\bf n}_2 
- \left( {\bf J}_{2R} + {\bf J}_{2L} \right) 
\cdot {\bf n}_1  \right]   
\nonumber \\
& & \ \ \ + ~ \frac{J_2 M}{a^{1/2}} \sum_{m=1}^{\infty} 
\frac{a^{2m}}{(2m)!}  \left[ \left( {\bf J}_{1R} 
+ {\bf J}_{1L} \right) \cdot \partial_x^{2m} {\bf n}_2  \right. 
\nonumber \\ 
& & \left. \hspace{1.25in}    
- \left( {\bf J}_{2R} + {\bf J}_{2L} \right) \cdot 
\partial_x^{2m} {\bf n}_1  \right]  \Big\}  \,.   \nonumber  
\end{eqnarray}  
Interestingly, even though the term ($\epsilon_1 - \epsilon_2$) is
allowed, its coefficient is zero.
However, the relevant  
$[( {\bf J}_{1R} + {\bf J}_{1L} ) \cdot {\bf n}_2 - ( 
{\bf J}_{2R} + {\bf J}_{2L} ) \cdot {\bf n}_1 ]$
is always present; as mentioned above, it generates both
the ${\bf n}_1 \cdot {\bf n}_2$ and $\epsilon_1 \epsilon_2$ 
terms.
Hence, for this model the irrelevant operators give rise
only to small quantitative corrections; the physics that
can occur is already contained in the relevant operators
already present.
Therefore, to leading order we
deduce the values of the parameters in 
\eqref{effectivediagonal} to be
\begin{eqnarray}
& & \lambda = \frac{J_{\bot} + J_2}{(2\pi)^2}  \,, \ \ \
g = \frac{(J_{\bot} - J_2) M^2}{(2\pi)^2}  \,, \ \ \  
\beta = 0 \,,   
\nonumber   \\   & &  \hspace{.85in}
\kappa = \frac{J_2 M}{(2\pi)^2} 
\,, \ \ \   \alpha = 0  \,.
\end{eqnarray}

As mentioned above, the zigzag ladder is expected to have
a fractionalized phase when $g \rightarrow 0\,$; one can 
expect a subtle competition between 
$\epsilon_1 \epsilon_2$ and
(${\bf J}_{1R} \cdot {\bf J}_{2L} + {\bf J}_{1L} \cdot {\bf J}_{2R}$)
in the cross-coupled ladder. For this ladder model, as 
$\epsilon_1 \epsilon_2$ is generated by the relevant 
$[( {\bf J}_{1R} + {\bf J}_{1L} ) \cdot {\bf n}_2 - ( 
{\bf J}_{2R} + {\bf J}_{2L} ) \cdot {\bf n}_1 ]$
term, it is reasonable to expect a regime of parameter space 
where $\epsilon_1 \epsilon_2$ flows to strong coupling first
and the system dimerizes.

%%%%%%%%%%%%%%%%%%%%%%%%%%%%%%%%%%%%%%%%%
%%%%%%%%%%%%%%%%%%%%%%%%%%%%%%%%%%%%%%%%%%

\section{Numerical Results and Discussion}

In this section, we present our numerical results for the phase 
diagrams of the models introduced in Sec.~II; we discuss them in 
light of the analytical results presented in the previous sections. 

%%%%%%%%%%%%%%%%%%%%%%%%%%%%%%%%%%%%%%%%

\subsection{Numerical Method}

The numerical calculations have been performed on finite ladders with 
open boundary condition (OBC) using the DMRG algorithm\cite{white} 
with the dynamic block-state selection (DBSS) 
approach.\cite{legeza02,legeza03}  
We have set the threshold value of the quantum information loss $\chi$ 
to $10^{-8}$ and the minimum number of block states $M_{\rm min}$ to 
$64$.  All relevant eigenstates have been targeted independently using 
four to six DMRG sweeps until the entropy sum rule has been satisfied.  
The accuracy of the Davidson diagonalization routine has been set to 
$10^{-7}$. 

Recently, it has been shown that quantum phase transitions (QPTs) can 
be conveniently studied by calculating some measure of 
entanglement.\cite{wootters,zanardi,osborne,osterloh,gu,vidal,gu2,wu,yang,legeza_qpt,deng}
In particular, the von Neumann entropy of a block containing a finite 
number of neighboring sites often gives a clear indication of a QPT, 
as anomalies appear in these quantities at the transition: the 
entropy exhibits a jump at a first-order transition or develops a cusp 
(with increasing $N$) at a continuous transition. 
It should be noted, however, that depending on how a first-order 
transition is realized, one might have difficulty distinguishing it 
from a continuous transition. More specifically, if the two levels 
corresponding to the different ground states are already orthogonal 
in a finite-sized system, the entropy of a block will exhibit a jump 
when the two levels cross (in a finite-sized system).
However, if the two levels are not orthogonal in a finite-sized system, 
the wave function, energy, and consequently the entropy of a block will 
vary continuously in any finite-sized system.  In this case, the level 
crossing develops only asymptotically, and the jump in the entropy 
appears only in the $N \rightarrow \infty$ limit.

In this work, we consider (i) $s_l$, the entropy of the $l^{\rm th}$ rung, 
(ii) $s_{l,l+1}$, the two-rung entropy of the neighboring $l^{\rm th}$ and 
$(l+1)^{\rm st}$ rungs, (iii) $s_{l,l+1}^{(i)}$, the two-site entropy of 
the spins ${\bf S}^{(i)}_l$ and ${\bf S}^{(i)}_{l+1}$ on chain $i$,
and also (iv) $s_N(l)$, the entropy of a block formed by the left $l$ 
rungs of a ladder with $N$ rungs.
To avoid end effects, we compute $s_l$, $s_{l,l+1}$, and $s_{l,l+1}^{(i)}$ 
in the middle of the ladder, for $l=N/2$ or $l=N/2+1$.
As discussed above, one of our primary interests is to identify
dimerized phases which may arise and the concomitant breaking of 
translational symmetry.  
The appearance of a columnar dimerized phase can be detected by 
considering the difference of two-rung entropies
\begin{equation}
D_s = s_{l+1,l+2}-s_{l,l+1}  \qquad   l=N/2 \,.
\label{eq:d_s}
\end{equation}  
Alternatively, taking the block entropy $s_N(l)$ of the left $l$ rungs, 
one can consider
\begin{equation}
\tilde{D}_s = s(l)-s(l+1)  \qquad   l=N/2 \,,
\end{equation}   
which tells how the block entropy of the left half changes when an extra 
rung is added. 
The appearance of staggered dimerization can be detected by considering
the difference of two-site entropies on the two chains
\begin{equation}
P_s = \big(s_{l+1,l+2}^{(1)}-s_{l,l+1}^{(1)}) 
 - (s_{l+1,l+2}^{(2)}-s_{l,l+1}^{(2)}\big)   \quad   l=N/2 \,. 
\label{eq:p_s}
\end{equation} 

Further information about the phases that arise can be deduced by 
studying the length dependence of 
$s_N(l)$.\cite{vidal_sl,korepin_sl,affleck,cardy,laflorence} 
For noncritical, gapped models, this quantity saturates to a finite value 
when $l$ is far from the boundaries, while for critical systems
\begin{equation}
s_N(l) = \frac{c}{6}\ln \left[ \frac{2N}{\pi} \sin \left( \frac{\pi l}{N}
\right) \right] + g  \,,
\label{eq:cardy}
\end{equation}
where $c$ is the central charge.\cite{holzhey,cardy} 
Moreover, if the system's ground state is spatially inhomogeneous, 
oscillations appear in $s_N(l)$.\cite{legeza_incomm} 
Hence, the Fourier spectrum 
\begin{equation}
 \tilde{s}(k) = \frac{1}{N}\sum_{l=0}^N e^{- i k l}s_N(l)
\label{eq:sq}
\end{equation}
carries information\cite{legeza_incomm} about the spatial inhomogeneity:
if the amplitude of a peak at a nonzero wave number $k^{\ast}$ remains 
finite in the thermodynamic limit, this indicates a periodic spatial 
modulation of the ground state with wavelength $\lambda=2\pi/k^{\ast}$.

In what follows, we will often need to know the large-$N$ behavior of
various quantities. For any quantity $A$, the finite-size-scaling Ansatz
\begin{equation}
A(N) = A_0 + a/{N^\beta}
\label{eq:scale_obc}
\end{equation}
has been used, where $A_0$, $a$, $\beta$ are free parameters determined 
by a least-squares fitting procedure.

%%%%%%%%%%%%%%%%%%%%%%%%%%%%%%%%%%%%%%%%%

\subsection{Results and Discussion}

\subsubsection{Zigzag ladder}  

We first present results for the zigzag ladder.  Since the phase 
diagram and properties of this model are well known, it serves as 
a test case for the other models.
The $J_{\bot}$, $J_2$ parameter space was explored by calculating 
various entropy functions for $J_{\bot}$ and $J_2$ satisfying
\begin{equation}
J_{\bot} + J_2 = C J_{\|}
\label{eq:j1j2}
\end{equation}
for several values of $C$. In the numerical calculations the energy scale was set
by taking $J_{\|}$ to be unity.

\begin{figure}[htb] 
\epsfxsize=8cm \epsfbox{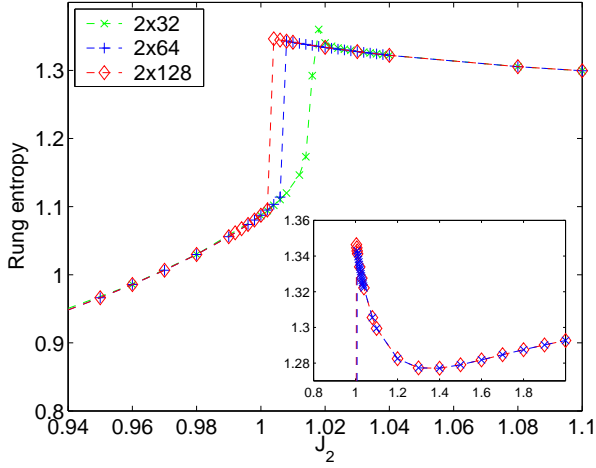} 
\caption{(Color online) Single-rung entropy as a function of $J_2$ for 
$J_{\bot} + J_2 = 2J_{\|}$ with $J_{\|} = 1$ for three different lengths of the ladder.} 
\label{fig:zigzag_path2} 
\end{figure}

We first show results obtained for $C=2$.
As seen in Fig.~\ref{fig:zigzag_path2}, $s_l$ in the middle of 
the ladder is discontinuous at $J_{\bot} = J_2 = J_{\|}$, indicating 
a first-order transition. 
Besides the discontinuity, the inset of Fig.~\ref{fig:zigzag_path2} 
shows a minimum in the entropy at $J_{2} \simeq 1.36 J_{\|}$, 
indicating a possible second transition.  Moreover, there is also 
a minimum in $\tilde s(k)$ at an incommensurate $k^{\ast}$ in the 
region $0.5 < J_2/J_{\|} < 1.36$. However, on physical grounds, there 
is no reason to expect a second transition in the zigzag ladder since 
any deviation from the $J_{\bot} = J_2$ line drives the system into 
the rung-singlet or Haldane phase. Indeed, this minimum is 
a finite-size effect --- it originates from the two end spins of the 
Haldane phase.  To substantiate this, notice that there is no sign 
of this minimum in the rung-singlet phase, while we know that the 
phase boundaries must be symmetric under the interchange of $J_{\bot}$ 
and $J_2$. Furthermore, by attaching spin-1/2's to the ends of the 
ladder with a strong anitferromagnetic coupling, the end spins can 
be eliminated.\cite{white_huse} When the calculations are repeated 
with these extra spin-1/2's attached to the ends, the minimum in 
$\tilde s(k)$ at an incommensurate $k^{\ast}$ disappears.

\begin{figure}[htb] 
\epsfxsize=8cm \epsfbox{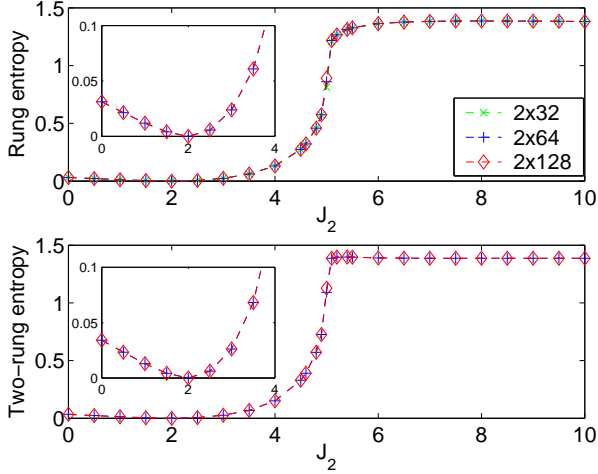} 
\caption{(Color online) Single-rung and block entropies for $J_{\bot} + J_2 =10J_{\|}$, as a 
function of $J_2$ for ladders with 32, 64 and 128 rungs. The insets show 
the behavior near $J_2 = 2J_{\|}$, where the entropies vanish.} 
\label{fig:zigzag_path10} 
\end{figure}

Performing the calculations for couplings satisfying \eqref{eq:j1j2} with
other values of $C$, a behavior similar to Fig.~\ref{fig:zigzag_path2} was 
observed whenever the line $J_{\bot} = J_2$ was crossed at 
$J_{\bot} < J_{\|}/0.241$. That indicates that the system 
undergoes a first-order transition (along the 
line $J_{\bot} = J_2$) for $J_{\bot} < J_{\|}/0.241$.
Different behaviors in the entropies were observed, however, for 
stronger interchain couplings.  Fig.~\ref{fig:zigzag_path10} shows $s_l$ 
(top panel) and $s_{l,l+1}$ (bottom panel) in the middle of the ladder  
for $J_{\bot} + J_2 = 10J_{\|}$.  
In the rung-singlet phase $s_l$ is small, as the spins on a rung are 
predominantly in a singlet state; in the Haldane-like phase $s_l$ is 
close to $\ln 4 \simeq 1.386$, since the valence bonds are formed 
predominantly between neighboring rungs. Notice, however, that $s_l$ 
no longer has a jump at $J_{\bot} = J_2 = CJ_{\|}/2$ 
(as it did in Fig.~\ref{fig:zigzag_path2}); rather, $s_l$ and $s_{l,l+1}$ 
have a discontinuity in their slope. 
When $s_N(l)$ was computed at the point $J_{\bot} = J_2$, it was found 
that it could be fit well with the form given in (\ref{eq:cardy}) with $c=1$. 
These results are consistent with a continuous transition between the 
rung-singlet and Haldane-like phases in this regime. They are consistent 
with the fact that, if the zigzag ladder is written as a single chain with 
nearest- and next-nearest-neighbor couplings, the chain is critical in 
this regime with the low-energy physics being described by 
\eqref{WZWhamiltonian}.

The insets of Fig.~\ref{fig:zigzag_path10} show $s_l$ and $s_{l,l+1}$
in a region about $J_2 = 2J_{\|}$; we see that the entropy functions vanish 
at $J_2 = 2J_{\|}$ for any length of the ladder.  This is because the exact
ground state is a product of rung singlets [see Eq.~\eqref{simpleexact}] for 
$J_2 = 2J_{\|}$.  In fact, the entropies vanishes along the entire line 
$J_2 = 2 J_{\|}$ when $J_{\bot} \geq 2J_{\|}$, as \eqref{simpleexact} is the 
exact ground state.  
Furthermore, computing the entropy of the two spins coupled along the
diagonal by $J_2$, ${\bf S}^{(1)}_l $ and ${\bf S}^{(2)}_{l+1}$, this
entropy was found to vanish for $J_{\bot} = 2 J_{\|}$ when $J_2 \geq 2J_{\|}$.
This is because \eqref{simpleexact-2} is the exact ground state 
along this line.  
More generally, due to the symmetry of the model, one obtains the 
same results presented in Figs.~\ref{fig:zigzag_path2} and 
\ref{fig:zigzag_path10} if $J_{\bot}$ and $J_{2}$ are interchanged;
instead of considering rung entropies, one must consider the entropies 
of diagonally coupled spins.

\begin{figure}[htb] 
\epsfxsize=8cm \epsfbox{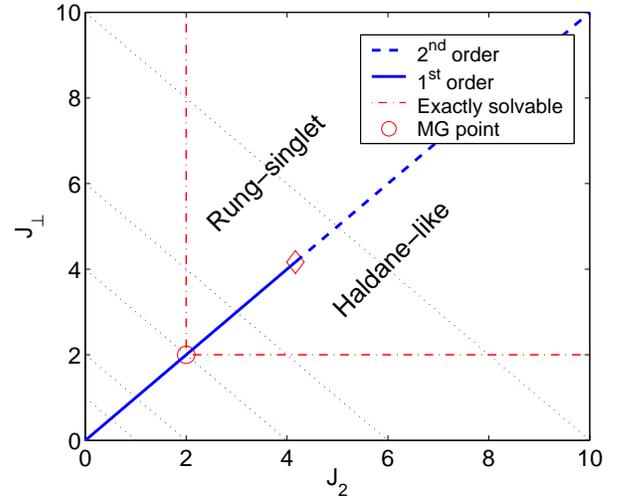} 
\caption{(Color online) Phase diagram of the zigzag ladder. The first- and second-order
transition between the rung-singlet and Haldane phases
are denoted by solid and dashed lines, respectively. The ground state along the
dashed-dotted line is an exact rung-singlet or diagonal-singlet valence-bond state.
The dotted lines indicate the parameter values wher the DMRG calculation were done.} 
\label{fig:zigzag_phase} 
\end{figure} 

Putting together these results, we arrive at the phase diagram in 
Fig.~\ref{fig:zigzag_phase}. The transition line at $J_{\bot}=J_2$ 
is of first order in the weak-coupling limit, which becomes of 
second order at strong couplings. 
The entire region above (below) the line $J_{\bot} = J_2$ is continuously 
related to the exactly solvable line $J_2=2J_{\|}$ ($J_{\bot}=2J_{\|}$) and, hence, is 
in the rung-singlet (Haldane) phase.  
In terms of the effective low-energy Hamiltonian discussed in the 
previous section, the physics in these phases is determined by the 
${\bf n}_1 \cdot {\bf n}_2$ term; in the rung-singlet (Haldane) 
phase, its coefficient is positive (negative). As discussed in Sec.~IV, 
the rung-singlet (Haldane) phase has $Q={\text{even}}$ ($Q={\text{odd}}$) topological 
order, and the elementary excitations are gapped magnons with spinons 
being confined.  Along the $J_{\bot} = J_2$ line, the two topologically distinct ground states
become degenerate, spinons are deconfined, being
domain walls between the two topologically distinct (and energetically
degenerate) ground states.
As mentioned in the previous section, at weak coupling the physics is
determined by the 
(${\bf J}_{1R} \cdot {\bf J}_{2L} + {\bf J}_{2R} \cdot {\bf J}_{1L}$) 
term (with the ${\bf n}_1 \cdot  \partial_x {\bf n}_2$  and 
$\epsilon_1 \partial_x \epsilon_2$ terms giving rise to 
incommensuration).
Looking at Eq.~\eqref{lowenergyzigzag}, this fractionalized phase occurs because the 
geometry of the zigzag ladder fine tunes away an infinite number of 
operators which could cause dimerization.

%%%%%%%%%%%%%%%%%%%%%%%%%%%%%%%%%%%%%%%%%%%%

\subsubsection{Cross-Coupled Ladder}  

\begin{figure}[htb] 
\epsfxsize=8cm \epsfbox{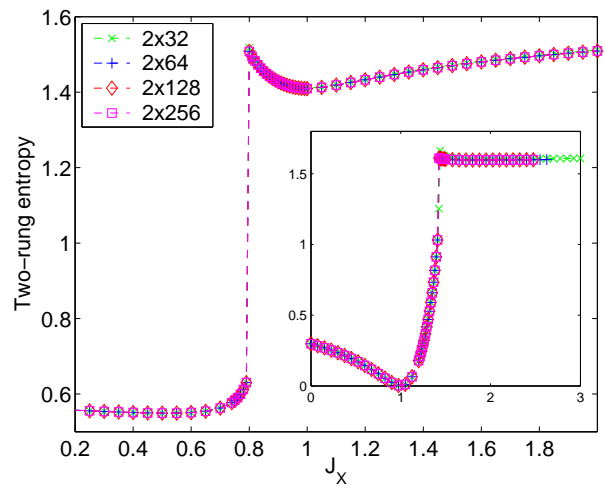} 
\caption{(Color online) Two-rung entropy of the cross-coupled ladder as a function of 
$J_{\text{X}}$ for $J_{\bot}=2J_{\|} - J_{\text{X}}$ for various system 
sizes. The dashed line is a guide to the eyes.  The inset shows the same 
but for the $J_{\bot} = 3J_{\|} - J_{\text{X}}$.} 
\label{fig:composite_s2_path2} 
\end{figure}

As with the zigzag ladder, we calculated $s_l$ and $s_{l,l+1}$ for 
$J_{\bot}$ and $J_{\text{X}}$ satisfying \eqref{eq:j1j2} with various 
values of $C$. 
Fig.~\ref{fig:composite_s2_path2} shows results obtained for intermediate 
values of the interchain couplings, $C=2$ and $C=3$.  Here, we see the 
entropy functions display a finite jump, indicating a first-order transition.
The inset shows that $s_l$ vanishes (independent of the ladder's length) 
at $J_{\text{X}} = J_{\|}$.  This occurs because \eqref{simpleexact} is 
the exact ground state for $J_{\text{X}} = J_{\|}$ with 
$J_{\bot} \geq 2J_{\|}$.
It was shown in Ref.~\onlinecite{weihong} that \eqref{simpleexact} is,
in fact, the ground state for $J_{\text{X}}=J_{\|}$ and 
$J_{\bot} \geq 1.401J_{\|}$.
Our numerical results are in agreement with this prediction.

Besides the finite jump at $J_{\text{X}} = 0.8$, 
Fig.~\ref{fig:composite_s2_path2} also shows that the two-site 
entropy possesses a minimum around $J_{\text{X}} \approx 1$; furthermore, 
weak incommensurate oscillations were found to appear in $s_N(l)$.  
Similar to the zigzag ladder, these are finite-size effects due 
to the end spins in the Haldane phase.  Indeed, when calculations 
are repeated for a system in which spin-1/2's are attached to the 
ends of the ladder (to freeze the end spins), the block entropy 
$s_N(l)$ saturates for shorter chains and the minimum in $\tilde{s}(k)$ 
at an incommensurate $k^{\ast}$ disappears. The amplitude of the 
remaining negative peak in $\tilde{s}(k)$ at $k^*=\pi$ was found to vanish 
in the thermodynamic limit.  Hence, no spatial inhomogeneity develops 
along the transition line.

\begin{figure}[htb] 
\epsfxsize=8cm \epsfbox{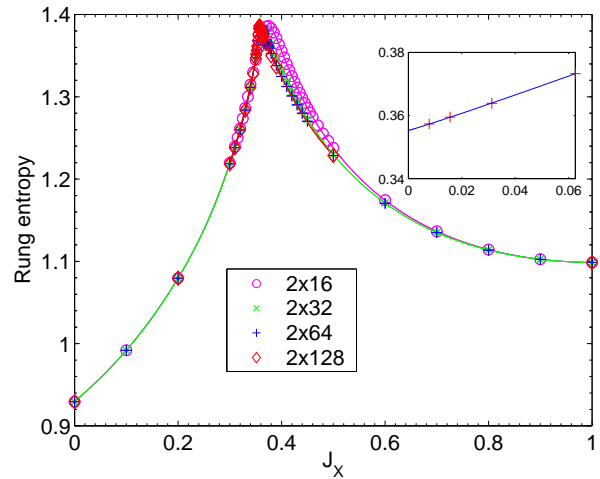} 
\caption{(Color online) One-rung entropy of the cross-coupled ladder model for 
$J_{\bot} + J_{\text{X}} = J_{\|}$, for several system sizes. 
The solid line is a polynomial fit.  
The inset shows the finite-size scaling of the position of the 
maximum of the entropy.}
\label{fig:composite_s1_path1} 
\end{figure} 

\begin{figure}[htb] 
\epsfxsize=8cm \epsfbox{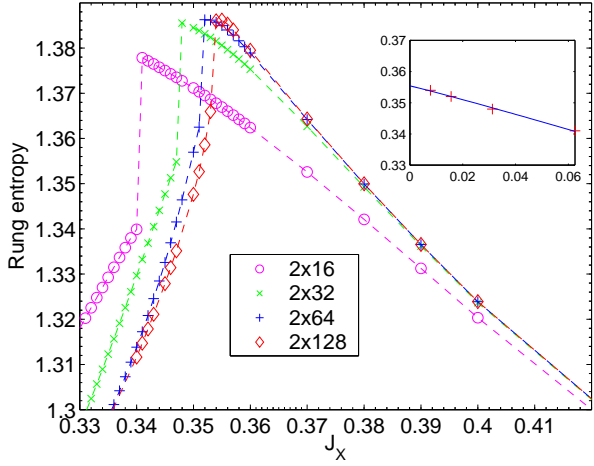} 
\caption{(Color online) Same as Fig.~\ref{fig:composite_s1_path1} but with spin-1/2 
particles attached to the two ends of the ladder
with strong antiferromagnetic couplings. 
The inset shows the finite-size scaling of the 
position of the jump in the entropy.}
\label{fig:composite_s1_path1_jend2} 
\end{figure}

The numerical results show different behavior when $C < 1.3$.  $s_l$ 
as a function of $J_{\text{X}}$ is shown in Fig.~\ref{fig:composite_s1_path1} 
for $C=1$. The ground-state wave function is continuous in the weak-coupling 
regime. As longer and longer ladders are considered, $s_l$ exhibits 
a sharper and sharper maximum (bounded from above by $\ln 4$), developing 
into a cusp at $J_{{\text{X}}c}/J_{\|} = 0.355(3)$.  Such behavior is 
suggestive of a continuous transition.  
Recent numerical works reported a continuous transition in the 
weak-coupling regime,\cite{wang,hung} in disagreement with analytic 
results\cite{gene} as well as previous numerical calculations.\cite{fath}
However, the analytic results are expected to be reliable in the 
weak-coupling regime.  As was checked by the DBSS procedure,
a reliable extrapolation of the gap requires calculations on longer ladders 
and keeping a significantly larger number of block states than was available in 
Ref.~\onlinecite{wang}. The same holds for the entropy. Moreover,
as was shown above, spurious effects can arise due to end spins.  
When the calculations were repeated by attaching spin-1/2's to the 
ends of the ladder; $s_l$ behaves somewhat differently.  
As shown in Fig.~\ref{fig:composite_s1_path1_jend2}, instead of an
abrupt change in slope, a jump now seems to develop; its position scales to 
the same value of $J_{{\text{X}}c}$ obtained in 
Fig.~\ref{fig:composite_s1_path1}. This behavior suggests that the 
transition is of first-order.
Based on these considerations, we believe the transition is, in fact, 
first-order. Although the singlet ground-state wave function is continuous in 
the weak-coupling regime for finite-sized systems, a crossing with the 
next singlet level (which we found to be at relatively high energy for weak couplings,
far from the exactly solvable line) may develop in the $N$$\rightarrow$$\infty$ 
limit, and the asymmetric cusp in Fig.~\ref{fig:composite_s1_path1} may 
develop into a jump.

\begin{figure}[htb] 
\epsfxsize=8cm \epsfbox{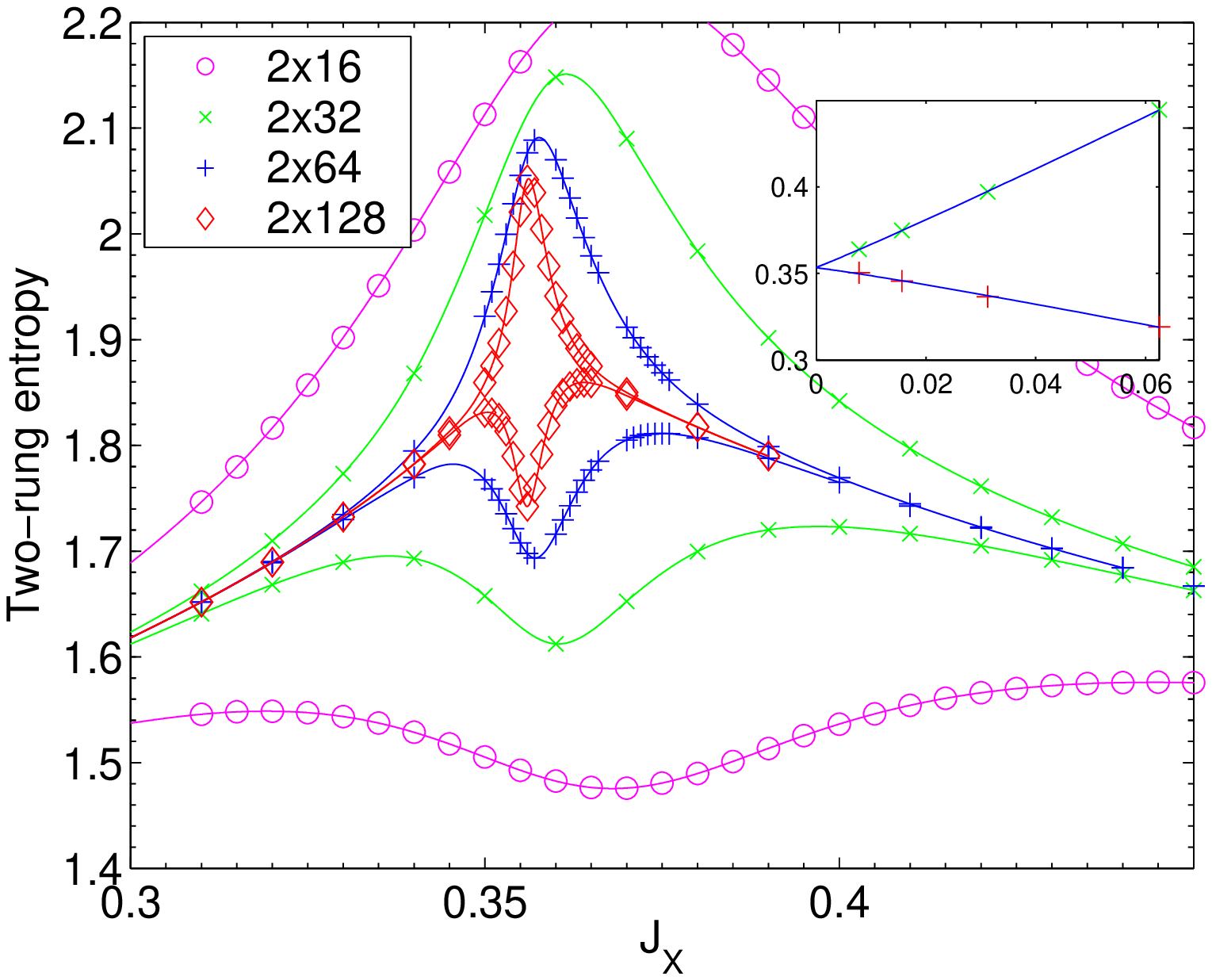} 
\caption{(Color online) Two-rung entropy of the cross-coupled ladder model for 
$J_{\bot} + J_{\text{X}} = J_{\|}$ for various system sizes. The solid line is a 
polynomial fit.  The inset shows the finite-size scaling of the 
positions of the peaks of the entropy.}
\label{fig:composite_s2_path1} 
\end{figure}

A further interesting feature of the model for weak interchain coupling 
is shown in Fig.~\ref{fig:composite_s2_path1}.  More specifically, the 
two-rung entropy $s_{l,l+1}$ measured for $l = N/2$ exhibits two well 
separated peaks at $J_{{\text{X}}c_1}(N)$ and $J_{{\text{X}}c_2}(N)$, but it exhibits a 
single peak for $l = N/2+1$. The difference $D_{\rm s}(N)$ is also finite 
in the region between $J_{{\text{X}}c_1}$ and $J_{{\text{X}}c_2}$.  
This could suggests the existence of a columnar dimer phase in a narrow 
range of couplings, as predicted in Ref.~\onlinecite{oleg}. However, the two 
peaks merge in the $N \rightarrow \infty$ limit and the width of the putative 
dimer phase shrinks to zero. As shown in the inset of 
Fig.~\ref{fig:composite_s2_path1}, a finite-size scaling analysis gives the 
same critical value as the one-rung entropy. The same behavior was
observed in the calculation along the line $J_{\text{X}} = 0.2J_{\|}$
confirming the findings of Ref.~\onlinecite{hung}.

We have also computed $s_N(l)$ and its Fourier transform $\tilde{s}(k)$. 
$|\tilde{s}(k)|$ was found to have an extra peak at $k^{\ast} = \pi$, 
besides the one at $k =0$.  However, $|\tilde s(k^*=\pi)|$ was found to 
vanish in the large-$N$ limit, indicating that the ground state is always 
spatially homogeneous.  Similar behavior was found for other values of 
$J_{\bot}$ and $J_{\text{X}}$ in the weak-coupling regime.  
Furthermore, we computed the staggered dimerization [see Eq.~\eqref{eq:p_s}]
and we found that it also vanishes in the $N \rightarrow 0$ limit.  
This provides strong evidence that, at least for weak interchain coupling, 
there is no intermediate columnar or staggered dimer phase between the 
rung-singlet and Haldane phases.

\begin{figure}[htb] 
\epsfxsize=8cm \epsfbox{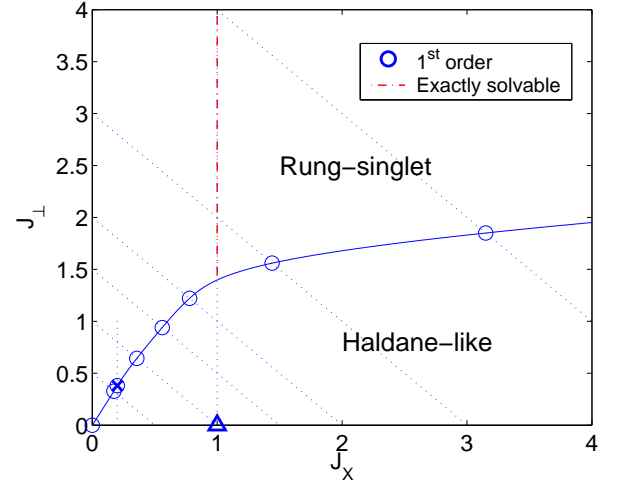} 
\caption{(Color online) Phase diagram of the cross-coupled ladder. 
The symbol $\times$ denotes the transition point calculated in Ref.~\onlinecite{hung}. 
$\triangle$ denotes the point where the ladder model is equivalent to a spin-1 chain. 
The ground state is an exact rung-singlet state along the dashed-dotted line.} 
\label{fig:composite_phase} 
\end{figure} 

Putting together these results, we obtain the phase diagram for the 
cross-coupled ladder, shown in Fig.~\ref{fig:composite_phase}.
The entire region above the transition line is continuously related 
to the exactly solvable line $J_{\text{X}}/J_{\|} =1$ and, hence, is in the rung-singlet 
phase.  
The entire region below the transition line is continuously related 
the point $J_{\text X} =1$, $J_{\bot}=0$ and, hence, is in the Haldane
phase.
As with the zigzag ladder, the physics in these phases is determined 
by the ${\bf n}_1 \cdot {\bf n}_2$ term in the low-energy Hamiltonian
--- these phases have topological order, and confined spinons. 
As discussed in the previous section, at the transition there is a subtle 
competition between the 
(${\bf J}_{1R} \cdot {\bf J}_{2L} + {\bf J}_{1L} \cdot {\bf J}_{2R}$) 
and $\epsilon_1 \epsilon_2$ terms, which give rise to a fractionalized
and dimerized phase, respectively.  However, the entire transition line 
appears to be of first-order, with no evidence for an intermediate 
dimerized phase being found.
Hence, our results suggest that a dimerized phase does not appear in this 
model.
In constructing the phase diagram, we made use of a duality 
relationship of the model. Similar to what was described for the 
diagonal ladder in Sec.~II, one can interchange the spins on every 
second rung: ${\bf S}_{2l}^1 \leftrightarrow {\bf S}_{2l}^2$.  
When this is done for this ladder model, another cross-coupled ladder 
is obtained but with $J_{\|}$ and $J_{\text{X}}$ interchanged.
This implies that energies and, in particular, energy gaps satisfy
\begin{equation}
E(J_{\|},J_{\bot},J_{\text{X}}) = E(J_{\text{X}},J_{\bot},J_{\|}) \,. 
\end{equation}
Scaling by $J_{\|}$, one obtains
\begin{equation}
E(J_{\bot}/J_{\|},J_{\text{X}}/J_{\|}) = 
(J_{\text{X}}/J_{\|})~ E(J_{\bot}/J_{\text{X}},J_{\|}/J_{\text{X}}) \,. 
\end{equation} 

%%%%%%%%%%%%%%%%%%%%%%%%%%%%%%%%%%%%%%%%%%%

\subsubsection{Diagonal Ladder}

As with the other models, we computed $s_l$ and $s_{l,l+1}$ for 
$J_{\bot}$ and $J_2$ satisfying \eqref{eq:j1j2} with various 
values of $C$.  Here, in contrast to the other two models, we 
find a dimerized phase intervening between the rung-singlet and 
Haldane phases.

\begin{figure}[htb] 
\epsfxsize=8cm \epsfbox{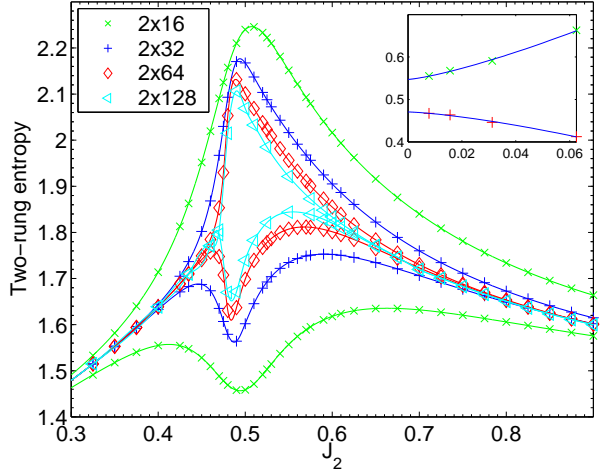} 
\caption{(Color online) Two-rung entropy of the diagonal ladder model for 
$J_{\bot} + J_2 = J_{\|}$ for various system sizes. The solid line is a 
polynomial fit.  The inset shows the finite-size scaling of the 
positions of the peaks of the entropy.}
\label{fig:diagonal_s2_path1} 
\end{figure} 

\begin{figure}[htb] 
\epsfxsize=8cm \epsfbox{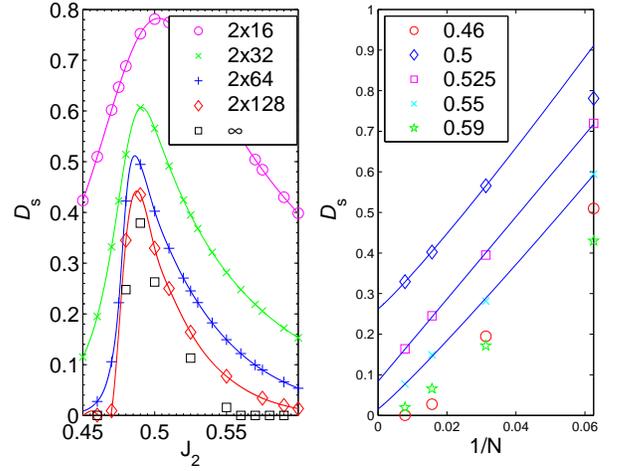} 
\caption{(Color online) The dimerization entropy of the diagonal ladder model for 
$J_{\bot} + J_2 = J_{\|}$ for various system sizes. The solid line is a 
polynomial fit.  The left panel shows the finite-size scaling of the 
$D_{\rm s}$ for various $J_2$ values.}
\label{fig:diagonal_ds_path1} 
\end{figure}

Fig.~\ref{fig:diagonal_s2_path1} shows $s_{l,l+1}$ for 
$J_{\bot} + J_2 = J_{\|}$.  While the behavior appears similar to 
Fig.~\ref{fig:composite_s2_path1}, the $N \rightarrow \infty$ 
limit is drastically different.  Indeed, while a single transition 
was found for the cross-coupled ladder, the inset of 
Fig.~\ref{fig:diagonal_s2_path1} shows the two peaks do not 
collapse in this model.  Hence, Fig.~\ref{fig:diagonal_s2_path1}
suggests the system undergoes two distinct transitions at 
$J_{2c_1}/J_{\|}=0.459$ and $J_{2c_2}/J_{\|}=0.563$, with a columnar dimerized 
phase between the two transition lines. 
To substantiate this, we have computed $s_N(l)$ and subsequently 
$|\tilde{s}(k^{\ast} = \pi)|$, the entropy difference between 
neighboring plaquettes $D_s$ (shown in 
Fig.~\ref{fig:diagonal_ds_path1}), and also the energy difference 
between neighboring plaquettes.  All of these quantities were found
to scale to a finite value in the region between 
$J_{2c_1}$ and $J_{2c_2}$. Similar behavior was found for other 
values of the parameters in the weak-coupling limit.
Hence, we conclude a columnar dimerized phase does, in fact, exist 
between the rung-singlet and Haldane phases in this model.

When the calculations were repeated for $J_{\bot} + J_2 > 2J_{\|}$, a 
single first-order transition was obtained. It is worth noting that 
due to the rather small values of the gap, the asymptotic behavior 
can be seen only for long ladders. As before, convergence can be 
accelerated in the Haldane phase by attaching spin-1/2's to the ends 
of the ladder (to pin the end spins).

\begin{figure}[htb]
\epsfxsize=8cm \epsfbox{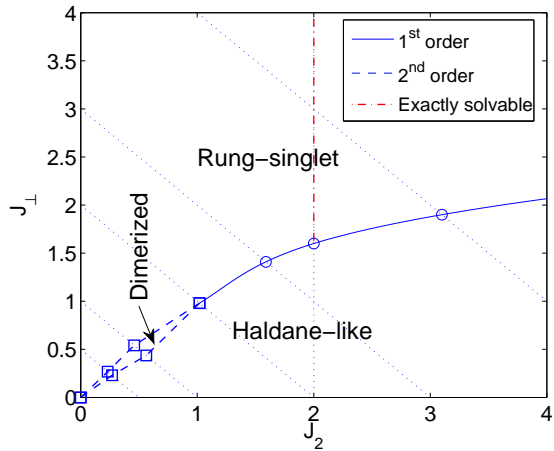}
\caption{(Color online) Phase diagram of the diagonal ladder.
The first and second-order phase transition points are
indicated by the circle and square symbols, respectively.
The exact rung-singlet ground state is indicated by a dashed-dotted line.}
\label{fig:diagonal_phase}
\end{figure}

Fig.~\ref{fig:diagonal_phase} summarizes our finding for the phase 
diagram of the diagonal ladder. 
The entire region above the transition line is continuously related 
to the exactly solvable line $J_2/J_{\|} =2$ and, hence, is in the 
rung-singlet phase.  
The entire region below the transition line is in the Haldane phase.
As with the previous two ladder models, these phases have topological 
order with confined spinons; the physics in these phases is determined 
by the ${\bf n}_1 \cdot {\bf n}_2$ term in the low-energy Hamiltonian.
However, contrary to what was found for the cross-coupled and zigzag 
ladders, from the discussion in the previous section, it is reasonable to expect
a regime in between the rung-singlet and Haldane phases where the 
diagonal ladder dimerizes (i.e., where the physics is determined by the
$\epsilon_1 \epsilon_2$ term).  This is, indeed, found to be the case
--- a narrow, but extended region is found where the ground state is 
dimerized. The first order transition is replaced in the weak coupling 
regime by two second-order transitions. 

%%%%%%%%%%%%%%%%%%%%%%%%%%%%%%%%%%%%%
%%%%%%%%%%%%%%%%%%%%%%%%%%%%%%%%%%%%%

\section{Concluding Remarks}

In this work we considered several frustrated spin ladder models, 
which are related to higher-dimensional models of current interest. 
In large regions of parameter space, these models have short-range 
RVB ground states with topological order; they 
are adiabatically related to the ground states of the rung-singlet 
or Haldane phase.
We investigated the role of frustrating interactions on the models, 
addressing in particular how the transition between phases with 
different topological order occurs.  
In the simplest case, a direct transition takes place along the 
line where the even- and odd-topology phases become degenerate.  
While the elementary excitations of the topologically ordered 
phases are gapped magnons, spinons become deconfined along the 
transition line and a fractionalized phase is obtained. 
Alternatively, the transition may occur in two steps, with an 
intermediate phase having broken translational symmetry.  In 
this case, spinons always remain confined.

An important observation from our analysis is the strong ``desire'' 
for broken-symmetry phases to arise at the transition between the 
rung-singlet and Haldane phases. Indeed, we saw that spin models 
typically give rise to an infinite number of operators that could 
cause dimerization.\cite{oleg} 
These operators arose in the diagonal ladder, and a dimerized
phase was seen to appear in the phase diagram.
These operators also arose in the cross-coupled ladder; nevertheless, in 
agreement with Ref.~\onlinecite{hung}, we found no evidence for 
dimerized phases in our numerics. The RG equations suggest that 
this occurs due to a subtle interplay/competition of quantum fluctuations.
Although no dimerized phase appears in the cross-coupled ladder
studied in this paper, in this delicate situation even small perturbations are 
likely to drive the cross-coupled model into dimerized phases as shown in 
Ref.~\onlinecite{vekua}. The zigzag ladder was an exception --- the 
model's geometry fine-tunes away the infinite number of operators which 
could cause dimerization.  

We believe these results give an outlook into the physics of
higher-dimensional systems.
In particular, our results show the importance of a system's
geometry in achieving the necessary liquidity for fractionalized
excitations to occur.
Hence, our results illustrate why fractionalized phases are hard
to come by --- fractionalized phases are delicate objects, 
requiring some level of fine-tuning.
Even on lattices where the necessary fine-tuning occurs (such
as the triangular lattice), small perturbations due to, e.g.,
spin-phonon coupling\cite{nt} or ring exchanges are likely to 
drive the system into a dimerized phase.

\section*{Acknowledgements}

EHK gratefully acknowledges the warm hospitality of the 
Research Institute for Solid State Physics and Optics 
(Budapest, Hungary), where parts of this work were performed.  
This work was supported by the NSERC of Canada (EHK), 
a SHARCNET Research Chair (EHK), a SHARCNET Senior Visiting
Scholar Award (EHK and JS),
and the Hungarian Research Fund OTKA 
Grant Nos. K 68340, F 46356,  and NF 61726 (OL and JS).

%%%%%%%%%%%%%%%%%%%%%%%%%%%%%%%%%%%%%%%%%
%%%%%%%%%%%%%%%%%%%%%%%%%%%%%%%%%%%%%%%

%%%%%%%%%%%%%%%%%%%%%%%%%%%%%%%%%%%%%
%%%%%%%%%%%%%%%%%%%%%%%%%%%%%%%%%%%%%%%%

\end{document}